\documentclass[11pt,a4paper]{article}
\pdfoutput=1
\usepackage{jheppub}
\usepackage{amsfonts}
\usepackage{bbm}
\usepackage{graphicx}
\usepackage{caption}
\usepackage{subcaption}
\usepackage{bigints}
\usepackage[normalem]{ulem}
\usepackage{slashed}
\usepackage{mathtools}
\DeclarePairedDelimiter{\ceil}{\lceil}{\rceil}

\def\a{\alpha}
\def\b{\beta}
\def\c{\gamma} \def\g{\gamma}

\def\e{\epsilon}

\def\i{\iota}

\def\l{\lambda}    
\def\m{\mu}
\def\n{\nu}

\def\th{\theta}                  
\def\r{\rho}                                     
                                  
\def\t{\tau}

\newcommand{\nn}{\nonumber}

\def\e{\epsilon}

\def\i{\mathrm{i}}

\def\p{\partial}

\def\bea{\begin{eqnarray}}
\def\eea{\end{eqnarray}}
\def\be{\begin{equation}}
\def\ee{\end{equation}}
\def\ba{\begin{align}}
\def\ea{\end{align}}
\def\ch {\cosh}
\def\sh{\sinh}

\newcommand{\bem}{\begin{pmatrix}}
\newcommand{\eem}{\end{pmatrix}}

\def\={\;  = \;}
\def\+{\, + \,}

\def\wt{\widetilde}

\def\bar{\overline}

\def\rt2{\sqrt{2}}

\title{Boundary Conditions and Localization  on  $AdS$: Part 1}

\author[a]{\small{Justin R.  David},}
\author[b]{\small{Edi Gava},}
\author[c]{\small{Rajesh Kumar Gupta},}
\author[d]{\small{Kumar Narain}}
\affiliation[a]{\small{Centre for High Energy Physics, Indian Institute of Science,\\
C. V. Raman Avenue, Bangalore 560012, India.}}
\affiliation[b]{\small{INFN, sezione di Trieste, Italy}}
\affiliation[c]{\small{Department of Mathematics, King's College London, The Strand, London WC2R 2LS, UK}}
\affiliation[d]{\small{ICTP, Strada Costiera 11, 34151 Trieste, Italy}}

\emailAdd{ justin@iisc.ac.in}
\emailAdd{gava@ictp.it}
\emailAdd{rajesh.gupta@kcl.ac.uk}
\emailAdd{narain@ictp.it}
\abstract{ We study the role of boundary conditions on the 
one loop partition function 
the ${\cal N}=2$  chiral multiplet   of R-charge $\Delta$ 
on $AdS_2\times S^1$. The chiral multiplet is coupled   to a background 
vector multiplet which preserves supersymmetry. 
We implement normalizable boundary conditions in $AdS_2$  and 
develop the Green's function method to obtain the one loop determinant.
We evaluate the one loop determinant  for two different actions: the standard 
action and the $Q$-exact deformed positive definite  action used for localization.  
We show that if there exists an integer $n$ in the interval  
$D: (  \frac{\Delta-1}{2L}, \frac{\Delta}{2L} )$, where $L$ being the ratio of radius of $AdS_2$ to that of $S^1$, then the one loop 
determinants  obtained for the two actions differ.  It is in this situation 
that fields which  obey normalizable boundary conditions do not 
obey supersymmetric boundary conditions.  However if there are 
no integers in $D$, then fields which obey normalizable boundary conditions
also obey supersymmetric boundary conditions and the one loop determinants
of the two actions precisely agree. 
We also show that  it is only in the latter  situation that the one loop determinant 
obtained by evaluating the index of the $D_{10}$ operator associated with the 
localizing action agrees with the  one loop determinant obtained 
using Green's function method. 
}

\begin{document}

%

\maketitle

\section{Introduction}
Supersymmetric localization methods  which was first introduced in 
\cite{Witten:1992xu}  and later developed in \cite{Nekrasov:2003af,Nekrasov:2003rj, Pestun:2007rz} 
enable the exact evaluation of 
observables in supersymmetric quantum field theories.  
See \cite{Pestun:2016zxk} for 
 a recent  review and a list of references. 
Some of the exact computations have provided highly non-trivial checks of 
the AdS/CFT correspondence \cite{Marino:2009jd,Benini:2015eyy,Benini:2016rke}. 
Localization relies on identifying a fermionic symmetry $Q$,   upto boundary terms
of the Lagrangian 
 and an addition of a  localizing term which is $Q$ exact up to boundary terms. 
 Therefore  supersymmetric theories on 
 compact spaces without boundaries serve as the canonical examples in which 
 the method of localization has been applied. 
 
 Extension of  localization methods to supersymmetric theories defined on 
 non-compact spaces presents a considerable challenge. 
 Naively we expect that we must ensure the following
 \begin{enumerate}
 \item Choose boundary conditions of all the fields involved such that 
 the   boundary terms that 
 result from the $Q$ variation of the 
 original action as well as the localizing term   can be neglected. 
 The boundary conditions must also be chosen so that the path integral is well defined. 
 The canonical method of ensuring this 
  is to implement normalizable boundary conditions. 
 \item The boundary conditions of both bosonic and fermionic fields chosen 
 must be consistent with supersymmetry. 
 \end{enumerate}
 It is a-priori not clear that fields which satisfy normalizable boundary conditions
 also obey supersymmetric boundary conditions. 
 To study these issues in detail it is best to focus on concrete examples of 
 a class of non-compact spaces.
 Supersymmetric theories on spaces of the form 
 $AdS_n\times S^m$  are examples which also have  good applications. 
 Localization of  ${\cal N}=2$ supergravity on $AdS_2\times S^2$ is relevant for 
 obtaining the entropy of extremal black holes in these theories 
 \cite{Dabholkar:2010uh,Dabholkar:2011ec,Gupta:2012cy,Dabholkar:2014ema,Murthy:2015yfa,Gupta:2015gga}. 
 Evaluating the supersymmetric partition function  of ${\cal N}=8$ supergravity 
 on $AdS_4$  is useful in the context of the holographic duality of this theory 
 with the ABJM theory \cite{Dabholkar:2014wpa}. Localization of supergravity on $AdS_{5}$ 
reduces to a Chern-Simons theory defined on an $AdS_{3}$ slice which describes a protected chiral algebra in $\mathcal N=4$ SYM theory in large $N$ limit \cite{Bonetti:2016nma}. Topologically twisted index on $AdS_2\times S^1$ of ABJM theory is related to the entropy of magnetically charged black holes in $AdS_{4}$ \cite{Cabo-Bizet:2017jsl}\,.  Apart from these applications, the study of the supersymmetric localization on non compact spaces itself is an interesting problem, see for instance \cite{Assel:2016pgi}. 
 
 As we have  briefly mentioned above,  since there are subtle issues involved 
 in the application of localization on non-compact spaces it is  important  to study 
 situations in which the results obtained by 
 applying localization  can be checked against another independent method. 
 In \cite{David:2016onq} localization of 
 ${\cal N}=2$   $U(N)$ Chern-Simons theory on $AdS_2\times S^1$
 was studied. 
 It was shown that it is possible to choose the  gauge field, the gaugino  and the 
 auxillary field  to  lie
 in the space of space of square integrable normalizable  wave functions 
  in $AdS_2$ and consistent
 with supersymmetry \footnote{The gauge fixing choice was such that  ghosts satisfied 
 supersymmetric boundary conditions, however they were not normalizable.}.
  Moreover the result for the supersymmetric partition function 
 for this theory on $AdS_2\times S^1$ agreed with that on $S^3$ which is expected 
 from conformal symmetry. 
 However since this theory is  also topological one might suspect that this agreement is
 the result of a coincidence and therefore it is important to examine the matter
 sector. 
  
 In this paper we study the localization of the chiral multiplet of R-charge $\Delta$ 
 on $AdS_2 \times S^1$ where the radius of $AdS_2$ is $L$  and the 
 radius of $S^1$  is normalized to unity. 
 The chiral multiplet is also coupled to the   background vector multiplet which 
 is supersymmetric and solves the saddle point equations of the localising 
 action of the vector multiplet. 
 The advantage of  studying the  matter sector is that its action is quadratic. 
 Therefore,  its partition function which is 
 one loop exact  can be obtained by using conventional methods and 
 compared against that obtained using localization. 
 Note that when the chiral multiplet is not  coupled with the vector, the theory is 
 free and the one loop determinants for the boson and the fermion in $AdS_2 \times S^1$
 for special values of $\Delta$ have been obtained earlier \cite{Klebanov:2011uf}
 using the eigen function method
 \footnote{In \cite{Klebanov:2011uf} thermal boundary conditions 
were chosen along $S^1$ for the fermions.  The method  can be adapted 
for fermions obeying  periodic boundary conditions along $S^1$.}.
 
We develop the Green's function approach of evaluating 
 the ratio of bosonic and fermionic one loop determinants in the chiral multiplet 
 coupled to the background vector.  
 We first consider the standard action for the chiral multiplet  given by \cite{Closset:2012ru}.
 We discuss in detail both  the normalizable and supersymmetric boundary conditions
 for fields of this theory. 
 We show that when  there exist no $n$ in the interval 
 \begin{equation} \label{interval1}
 D: \quad \Big( \frac{\Delta -1}{2L}, \frac{\Delta}{2L} \Big)\,,
 \end{equation}
solutions to equations of motions  which obey normalizable boundary conditions also obey
supersymmetric boundary conditions \footnote{In this paper, we will refer to this situation as 
normalizable boundary conditions are compatible with supersymmetric boundary conditions.}.
 We  show that the Green's function for both the boson and the fermion of the chiral multiplet 
 in the background of the vector multiplet is exactly solvable. 
 The Green's function can then be related to the variation of one loop determinants
  with respect to the 
 background value of the vector multiplet in the following way. 
 Let the background value of the vector multiplet be parametrised by $\alpha$. 
 The variation of the one loop  bosonic   determinant of  an operator   ${\cal D}_B(\alpha)$
 together with a fermionic determinant of an operator ${\cal D}_F(\alpha)$ 
  is 
 given by 
 \begin{equation}\label{green}
 \frac{\delta}{\delta\alpha} \ln Z_{\rm 1-loop}(\a) = {\rm Tr} [ G_F
 \frac{\delta}{\delta\alpha} {\cal D}_{F}(\alpha)  ] 
 -\frac{1}{2}  {\rm Tr} [ G_B
 \frac{\delta}{\delta\alpha} {\cal D}_{B}(\alpha)  ] \,.
 \end{equation}
 Then integrating with respect to  $\alpha$ enables the evaluation of the 
partition function.   The result is presented in (\ref{mainresult})\,.
 At $\alpha =0$
and special values of $\Delta, L$, the one loop determinants can be evaluated using 
the eigen function methods following \cite{Klebanov:2011uf} since the theory is free. 
As a check on the Green's function method we show the result for the one loop 
determinant at $L=1, 2$ and at $\alpha =0$ agrees with the eigen function 
method. 

Next we examine the $Q$-exact and positive definite action  which is used in 
localization 
 and again evaluate its 
one loop determinant using the Green's function method.  
We show that the one loop determinant of the standard action 
agrees with the $Q$-exact action only when there exist no integer $n$ in the domain 
$D$. As mentioned earlier it is in this situation that fields which obey normalizable 
boundary conditions  also satisfy supersymmetric boundary conditions. 
We also show  that  when there exists no integer $n$ in $D$, the one loop determinant entirely 
arises from boundary terms which include contributions from $r=0$ and $r\rightarrow\infty$ in $AdS_2$ and the result is independent of the 
details of the precise solutions to equations of motion of the action. 
Now  when  there exists an 
integer $n$ which lies in $D$, then the one loop determinant of the standard action 
differs from that of the $Q$-exact deformation. 
This shows  that the one loop determinant certainly is  not independent of the 
$Q$-exact deformation when  fields obeying normalizable boundary conditions 
do not satisfy supersymmetric boundary conditions.

 We also calculate  the one loop determinant of the $Q$-exact action  using the index 
 of the corresponding $D_{10}$ operator and implementing supersymmetric boundary 
 conditions. 
 We observe that when there exists no integer $n$ in the domain $D$ as well as $\frac{\Delta-1}{2L}$ is not an integer, then the 
 one loop determinant agrees with that obtained for this action as well as the standard action 
 using the Green's function method.  This is in accordance with the expectation that 
 when normalizable and supersymmetric boundary conditions are compatible, 
 methods relying on localization should yield the same answer as Green's function 
 approach implemented with normalizable boundary conditions. 
  
The organisation of this paper is as follows. 
In section  \ref{actions}  we discuss the supersymmetric transformations as well as the 
construction of the supersymmetric action for chiral multiplet on $AdS_2\times S^1$. 
In section \ref{regaction}
we introduce co-homological variables for the  fermions and write down the  Kaluza-Klein 
reduced actions  for the chiral multiplet in presence of the vector multiplet. 
We call this the standard  action. 
This is then repeated for the localising action of the chiral multiplet, we call this 
the $Q$-exact deformed action in section \ref{Sec:Q-exact.1}. 
Next  in section \ref{sec.BdyConds.1}
we  discuss the boundary conditions of the solutions to the 
equations of motion of these  Lagrangians. 
In section  \ref{1-loop}
we construct the Green's function for operators associated with 
the standard action of the  chiral multiplet as well as the $Q$-exact action. 
Here we make the main observation of the paper. 
We show that  if there exists an integer $n$ in the interval 
(\ref{interval1})
 the result for the one loop determinants for the regular action 
differs from that of the $Q$-exact action. 
We also observe here that once the interval 
(\ref{interval1}) admits an integer
 there are solutions which  obey normalizable boundary conditions, but do not satisfy 
 supersymmetric boundary condition.
 This is the reason that the one loop determinants of the 
$Q$-exact action does not agree with the regular action. 
In section \ref{indexd10}
we evaluate the index of the $D_{10}$ operator associated with the 
localising $Q$-exact action. Here we observe that it coincides with the result 
obtained using the  Green's function
method  for the standard as well as the $Q$-exact  action only when  
the interval (\ref{interval1}) 
does not admit an integer and $\frac{\Delta-1}{2L}$ is not an integer. 
In section \ref{conclusions} we present our conclusions and discuss generalisations. 
Appendix \ref{appendixa} contains the details regarding our conventions, killing spinors
and the classical  supersymmetric solution for  the vector multiplet. 
Appendix \ref{appendixb} contains the evaluation of one loop determinants of the 
free chiral multiplet  on $AdS_2\times S^1$using the eigen function method. 
Appendix \ref{appendixc}  tabulates a list of integrals involving products of 
hypergeometric functions which are used in our evaluation
of one loop determinants.

\section{Supersymmetry and actions on AdS$_{2}\times$S$^{1}$} \label{actions}

Before we start discussing the actions,   the preliminaries that we require are the 
supersymmetry transformations of the vector multiplet and a chiral multiplet 
coupled to a vector multiplet on $AdS_2\times S^1$. 
We take the ratio of the radius of $AdS_2$ to that of $S^1$ to be  $L$. 
The metric is given in (\ref{metric}). 
Let us begin with  the Euclidean supersymmetry 
 transformations of the fields in a vector multiplet. This is given by 
\bea \label{vecsusy}
&&Q\lambda=-\frac{i}{4}\epsilon \,G-\frac{i}{2}\epsilon^{\mu\nu\rho}\gamma_\rho F_{\mu\nu}\epsilon-i\gamma^\mu\epsilon\left(i\nabla_\mu\sigma-V_\mu\sigma\right)\,,\nn\\
&&Q\tilde\lambda=\frac{i}{4}\tilde\epsilon \,G-\frac{i}{2}\epsilon^{\mu\nu\rho}\gamma_\rho F_{\mu\nu}\tilde\epsilon+i\gamma^\mu\tilde\epsilon\left(i\nabla_\mu\sigma+V_\mu\sigma\right)\,,\nn\\
&&Qa_\mu=\frac{1}{2}\left(\epsilon\gamma_\mu\tilde\lambda+\tilde\epsilon\gamma_\mu\lambda\right)\,,\nn\\
&&Q\sigma=\frac{1}{2}\left(-\epsilon\tilde\lambda+\tilde\epsilon\lambda\right)\,,\nn\\
&&QG=-2i\left[\nabla_\mu\left(\epsilon\gamma^\mu\tilde\lambda-\tilde\epsilon\gamma^\mu\lambda\right)-i\left[\sigma,\epsilon\tilde\lambda+\tilde\epsilon\lambda\right]-iV_\mu\left(\epsilon\gamma^\mu\tilde\lambda+\tilde\epsilon\gamma^\mu\lambda\right)\right]\,.
\eea
where $\nabla_\mu$ is the covariant derivative  containing the Christoffel and the 
gauge  connection. 
The Killing spinors
 $\epsilon$ and $\tilde\epsilon$ on $AdS_2\times S^1$
 as well as the background $V_\mu$ is defined 
in (\ref{susy-backgrd}) \footnote{Please see appendix \ref{appendixa} 
for our notations and conventions.}.  
The supersymmetry transformations of the fields of the  chiral multiplet  coupled 
to an abelian vector multiplet are given by 
\bea
&&Q\phi=\epsilon\psi\,,\qquad Q\bar\phi=\tilde\epsilon\wt\psi\,,\nn\\
&&Q\psi = F(\t,r,\th)\epsilon+\Gamma^{\m}\tilde\epsilon\,  
D_{\m}\phi-iq\sigma\,\phi\,\tilde\epsilon\,,\nn\\
&&Q\wt\psi=\bar F(\t,r,\th)\tilde\epsilon+\Gamma^{\m}\epsilon\, 
D_{\m}\bar\phi-iq\sigma\,\bar\phi\,\epsilon\,,\nn\\
&&QF=D_{\m}(\tilde\epsilon\Gamma^{\m}\psi)+
iq\sigma\,\tilde\epsilon\psi-iq\phi\,\tilde\epsilon\wt\lambda\,,\nn\\
&&Q\bar F= D_{\m}(\epsilon\Gamma^{\m}\wt\psi)+iq\sigma\,\epsilon\wt\psi+
iq\bar\phi\,\epsilon\lambda\,.
\eea
Here $q$ refers to the charge of the  chiral multiplet. 
The fields $( \phi, \psi, F)$ has  R-charges  $(\Delta, \Delta -1, \Delta-2)$, while 
the fields of the anti-chiral multiplet 
 $(\bar \phi, \wt\psi, \bar F)$ has  R-charges $(-\Delta, -\Delta +1, -\Delta+2)$.
 The action of the  derivative $D_\mu$ are defined by 
 \begin{eqnarray}
 D_\mu \phi &=&  (\nabla_\mu - i \Delta A_\mu + i \frac{\Delta}{2} V_\mu ) \phi,  \\ \nonumber
 D_\mu \psi &=& ( \nabla_\mu - i (\Delta - 1) A_\mu + i \frac{\Delta}{2} V_\mu ) \psi, \\ \nonumber
 D_\mu \bar\phi &=& ( \nabla_\mu + i \Delta A_\mu - i \frac{\Delta}{2} V_\mu ) \bar \phi, 
 \\ \nonumber
 D_\mu \wt\psi &=& ( \nabla_\mu  + i ( \Delta -1) A_\mu - i \frac{\Delta}{2} V_\mu )\wt\psi
 \end{eqnarray}
 Note that $\nabla_\mu$ is the covariant derivative along with the gauge connection. 
It is convenient to define the variation separately with respect 
$\epsilon$ and $\tilde\epsilon$  as 
\bea
&&\delta_{\epsilon}\phi=\epsilon\psi\,,\quad \delta_{\tilde\epsilon}\phi=0,\quad 
\delta_{\epsilon}\bar\phi=0,\quad \delta_{\tilde\epsilon}\bar\phi=\tilde\epsilon\wt\psi\,,\nn\\
&&\delta_{\epsilon}\psi = F(\t,r,\th)\epsilon,\qquad \delta_{\tilde\epsilon\xi}\psi=
\Gamma^{\m}\tilde\epsilon\, D_{\m}\phi-iq\sigma\,\phi\,\tilde\epsilon\,,\nn\\
&&\delta_{\epsilon}\wt\psi=\Gamma^{\m}\epsilon\,  D_{\m}\bar\phi-iq\sigma\,\bar\phi\,\epsilon\,,\qquad \delta_{\tilde\epsilon}\wt\psi=\bar F(\t,r,\th)\tilde\epsilon\,,\nn\\
&&\delta_{\epsilon}F=0,\qquad \delta_{\tilde\epsilon}F=D_{\m}(\tilde\epsilon\Gamma^{\m}\psi)+iq\sigma\,\tilde\epsilon\psi-iq\phi\,\tilde\epsilon\wt\lambda\,,\nn\\
&&\delta_{\epsilon}\bar F= D_{\m}(\epsilon\Gamma^{\m}\wt\psi)+iq\sigma\,\epsilon\wt\psi+iq\bar\phi\,\epsilon\lambda\,,\qquad \delta_{\wt\epsilon}\bar F=0.
\eea
Then $Q$ is the sum given by 
$Q = \delta_{\epsilon} + \delta_{\wt\epsilon}$. 
The  action of $Q^2$ on all of the fields can be written compactly
as 
\be
Q^2=\mathcal L_K+\delta^{\text{gauge transf}}_\Lambda+\delta^{R-\text{symm}}_{\frac{1}{2L}}\,.
\ee
Here $ \mathcal L_K$ refers to the Lie derivative along the direction of the 
Killing vector
\begin{equation}
K^\mu = \tilde{\epsilon} \gamma^\mu \epsilon , \qquad 
K = \frac{\partial}{\partial \tau}  + \frac{1}{L} \frac{\partial}{\partial\theta}\,,
\end{equation}
and $\Lambda=\wt\epsilon\epsilon\,\sigma-K^{\rho}a_{\rho}$\,.
We now will need the vector multiplet background about which we will 
evaluate the one loop determinants of the chiral multiplet. 
We take this background to be given by 
\be \label{backvec}
a_\mu=0\,,\quad\sigma=\frac{i\alpha}{\cosh r}\,,\quad G=\frac{4i\alpha}{L\cosh^2r}\,.
\ee
Here $\alpha$ is a real constant which is matrix valued in the Lie algebra. 
We can chose it  to 
lie in the 
Cartan of the gauge group.
One can easily  verify that this background is invariant under the supersymmetric variation
given in (\ref{vecsusy}). 
This background is  also the classical solution of the equations of motion as well as minima to 
 the following localizing action 
of the vector multiplet. 
\bea\label{bosonLagrm}
QV_{loc\{\text{bosonic}\}}
&=&\int d^3x\sqrt{g}\text{Tr}\left[\frac{1}{4}F_{\mu\nu}F^{\mu\nu}-\frac{1}{2\cosh^2r}D_\mu(\cosh r\,\sigma)D^\mu(\cosh r\,\sigma) \right.  \nonumber \\
& & \left. \qquad\qquad \qquad\qquad-\frac{1}{32}\left(G-\frac{4\sigma}{L\cosh r}\right)^2\right]\,.
\eea

This localizing  action was used to obtain the  supersymmetric partition function 
of Chern-Simons theory on $AdS_2\times S^1$ in \cite{David:2016onq}. 

\subsection{The standard action  on AdS$_{2}\times$S$^{1}$} \label{regaction}

In this section we consider a chiral multiplet coupled to an abelian vector multiplet with charge $q$ on supersymmetric AdS$_{2}\times$S$^{1}$ background. Since in this paper we are restricting ourselves to only chiral multiplet, the abelian vector field, which couples to either gauge current or global current like flavor current, is restricted to the susy background \eqref{backvec}. The generalization to the case of chiral multiplet coupled to a non abelian gauge field with gauge group $G$ is straight forward. The supersymmetric action is given by
\bea\label{eq:susyS.1}
S&=&\int\,d^{3}x\sqrt{g}\Big[\mathcal D_{\mu}\bar\phi \mathcal D_{\mu}\phi+\Big(-\frac{1}{4}qG-\frac{\Delta}{4}R+\frac{1}{2}\left(\Delta-\frac{1}{2}\right)V^{2}-q^{2}\sigma^{2}\Big)\bar\phi\phi-F\bar F\nn\\&&+\wt\psi\slashed {\mathcal D}\psi+iq\sigma\,\wt\psi\psi+iq\bar\phi(\l\psi)-iq\phi(\wt\psi\wt\l)\Big]\,,
\eea
where
\bea
&&\mathcal D_{\mu}\phi=\p_{\mu}\phi-i\Delta\left(A_{\mu}-\frac{3}{2}V_{\mu}\right)\phi-i\left(\Delta-\frac{1}{2}\right)V_{\mu}\phi-iq a_{\mu}\phi=D_{\m}\phi+\frac{i}{2}V_{\m}\phi\,,\nn\\
&&{\mathcal D}_{\mu}\psi=\nabla_{\mu}\psi-i(\Delta-1)\Big(A_{\mu}-\frac{3}{2}V_{\mu}\Big)\psi-i\Big(\Delta-\frac{1}{2}\Big)V_{\mu}\psi-iqa_{\mu}\psi=D_{\m}\psi-\frac{i}{2}V_{\m}\psi\,.\nn\\
\eea
In the above action $\Delta$ is the R-charge of the chiral multiplet, $A_{\mu}$ and $V_{\mu}$ are supergravity background fields whose values are $A_{\tau}=V_{\tau}=\frac{1}{L}$ and $a_{\mu}$ is $U(1)$ gauge field and $R$ is the Ricci scalar. 
This action for the chiral multiplet
 was considered in \cite{Closset:2012ru} and we will refer it to as 
the `standard action'. The action \eqref{eq:susyS.1} is supersymmetric as well as $Q$-exact upto boundary terms
\be\label{eq:QexactS.1}
S=\int\,d^{3}x\sqrt{g}\frac{1}{\cosh r}[\delta_{\tilde\xi}\delta_{\xi}(\wt\psi\psi+\frac{i}{L\cosh r}\bar\phi\phi+2iq\sigma\bar\phi\phi)]+\text{boundary terms}\,.
\ee
 The boundary terms are given by 
\begin{eqnarray}\label{boundterm}
\text{boundary terms}&=&\int\,d^{3}x\sqrt{g}\,\nabla_{\m}\left[-\frac{1}{\cosh r}(\wt\e\wt\psi)(\e\g^{\m}\psi)-\frac{i}{2}V^{\m}\bar\phi\phi \right.  \\ \nonumber
& & \left. \qquad\qquad\qquad  -\frac{i}{\cosh r}\varepsilon^{\m\r\n}(\e\g_{\n}\wt\e)\bar\phi\,D_{\r}\phi+(\e\g^{\m}\wt\e)\frac{iq\sigma}{\cosh r}\bar\phi\phi\right]\,.
\end{eqnarray}
It will be pointed out 
in the section \ref{sec.BdyConds.1} that the above boundary terms vanish with both normalizable as 
well as 
supersymmetric boundary conditions.
Therefore, we can apply the technique of supersymmetric localization to the chiral multiplet on $AdS_{2}\times S^{1}$ to evaluate one loop determinants  of the fields in the chiral multiplet. 
Note that we will consider the action in (\ref{eq:susyS.1})  in the background of the 
vector multiplet given in (\ref{backvec})\,.

\noindent
{\bf Twisted variables}:  To proceed with the analysis it is convenient to 
 define the following twisted variables
\be\label{twistVar.1}
B(\tau,r,\th)=\wt\epsilon\psi,\quad \wt B(\tau,r,\th)=\epsilon\wt\psi,\quad C(\tau,r,\th)=\epsilon\psi,\quad \wt C(\tau,r,\th)=\wt\epsilon\wt\psi\,.
\ee
The map to the twisted variables is an one to one map and it can be inverted. 
In term of these variables the fermions $\psi$ and $\wt\psi$ are given as
\be\label{twist2}
\psi=\frac{1}{\wt\epsilon\epsilon}\Big(-\wt\epsilon\,C(\tau,r,\th)+\epsilon\,B(\tau,r,\th)\Big),\quad \wt\psi=\frac{1}{\wt\epsilon\epsilon}\Big(\epsilon\,\wt C(\tau,r,\th)-\wt\epsilon\,\wt B(\tau,r,\th)\Big)\,.
\ee

\noindent
{\bf Kaluza-Klein reduction:} 
In order to simplify the action, we decompose the fields into the Fourier modes labelled by $(n,p)$ 
along the $S^1$ and the angle direction of $AdS_2$ as
\bea\label{eq:FourierDecmp.1}
&&\phi(\tau,r,\th)=e^{i(n\,\tau+p\,\th)}f_{n,p}(r),\qquad\qquad \bar\phi(\tau,r,\th)=e^{-i(n\,\tau+p\,\th)}\bar f_{n,p}(r)\,,\nn\\
&&C(\tau,r,\th)=e^{i(n\,\tau+p\,\th)}c_{n,p}(r),\qquad\qquad \wt C(\tau,r,\th)=e^{-i(n\,\tau+p\,\th)}\wt c_{n,p}(r)\,,\nn\\
&&B(\tau,r,\th)=e^{i(n\,\tau+(p-1)\,\th)}b_{n,p}(r),\qquad\,\, \wt B(\tau,r,\th)=e^{-i(n\,\tau+(p-1)\,\th)}\wt b_{n,p}(r)\,,\nn\\
&&F(\tau,r,\th)=e^{i(n\,\tau+(p-1)\,\th)}F_{n,p}(r),\qquad \,\bar F(\tau,r,\th)=e^{-i(n\,\tau+(p-1)\,\th)}\bar F_{n,p}(r)\,.
\eea
In this case the bosonic part of the action (\ref{eq:susyS.1})  for a given $(n,p)$ becomes
\bea
S_{B,\{n,p\}}&=& \frac{1}{4}\int dr\,\sh r\Big[-4L^{2}F_{n,p}\bar F_{n,p}+\Big(4Ln+4L^{2}n^{2}-2\Delta-4Ln\Delta+\Delta^{2}+\frac{4p^{2}}{\sh^{2}r}\nn\\&&+\frac{4Lq\alpha(-i+Lq\alpha)}{\ch^{2}r}\Big)f_{n,p}(r)\bar f_{n,p}(r)+4\p_{r}f_{n,p}(r)\p_{r}\bar f_{n,p}(r)\Big]\,,
\eea
and similarly the fermionic part for a given $(n,p)$ is given by
\bea
S_{F,\{n,p\}}=-\frac{L}{4}\int dr &&\frac{1}{\ch^{2}r}\Big[\wt c_{n,p}(r)\Big\{b_{n,p}(r)(-4+2Ln+4p-\Delta+(-2Ln+\Delta)\ch2r)\nn\\&&+2\sh r\Big(i(-2+2Ln+2p-\Delta-2iLq\alpha)c_{n,p}(r)-2\ch r\,b_{n,p}'(r)\Big)\Big\}\nn\\&&+\wt b_{n,p}(r)\Big\{c_{n,p}(r)(2Ln+4p-\Delta+(-2Ln+\Delta)\ch2r)\\&&+2\sh r\Big(i(2Ln+2p-\Delta+2iLq\alpha)b_{n,p}(r)+2\ch r\,c_{n,p}'(r)\Big)\Big\}\Big]\,.\nn
\eea
Note that in these Kaluza-Klein reduced actions we have substituted the background 
vector multiplet in (\ref{backvec}). 
Since $F$ is an auxiliary field, its contribution to the one loop determinant
is trivial.  Hence forth we will drop it from the action. 

\noindent
{\bf Change of variables:}
We next change to the variable $z$ by defining $z=\tanh^{2}r$.  Now the origin of $AdS_2$ is 
at $z=0$, while the boundary is at $z =1$. 
In this variable the bosonic part of the action becomes  
\bea
S_{B,\{n,p\}}&=&\int^{1}_{0} dz\,\frac{1}{8z(1-z)^{3/2}}\Big[f_{n,p}(z)\bar f_{n,p}(z)\Big\{4p^{2}(1-z)+z\Big(\Delta(\Delta-2)+L(4n(1-\Delta)\nn\\
&&+4iq(z-1)\alpha)+4L^{2}(n^{2}-q^{2}(z-1)\alpha^{2})\Big)\Big\}+16z^{2}(1-z)^{2}f_{n,p}'(z)\bar f_{n,p}'(z)\Big]\,.\nn\\
\eea
In the above we have also included the contribution coming from change of integration measure (i.e. change from $r$ to $z$).\\
Now we vary the action w.r.t $\bar f_{n,p}$ to obtain the equation of motion for $f_{n,p}(z)$ which is\footnote{The differential operator is not hermitian for real $\a$. This is also true for the operator obtained for fermionic field in which case it is not anti-hermitian. We assume that $\a$ is imaginary for which it is hermitian in bosonic case and anti hermitian in fermionic case and at the end we analytically continue back to real $\a$. } 
\bea\label{eq:f.AdS2.1}
&&2z\sqrt{(1-z)}\,\p^{2}_{z}\,f_{n,p}(z)+\frac{(2-3z)}{\sqrt{1-z}}\,\p_{z}f_{n,p}(z)+\frac{1}{8z\sqrt{(1-z)^{3}}}\Big(-4p^{2}(1-z)+z(2-\Delta)\Delta\nn\\
&&+4Lz(n(-1+\Delta)+iq\a(1-z))-4L^{2}z(n^{2}+q^{2}\a^{2}(1-z))\Big)f_{n,p}(z)\=0\,.
\eea
Similarly the fermionic part of the action written in the coordinate $z$ is given as
\bea\label{eq;Sf.AdS2.1}
S_{F,\{n,p\}}&=&\int^{1}_{0} dz\,\frac{1}{2}\begin{pmatrix}\tilde b_{n,p}(z)&\tilde c_{n,p}(z)\end{pmatrix}\Big(-2iL\sqrt{z}\,\sigma_{2}\,\p_{z}+\frac{L(1-2p(1-z)+(-1+2Ln-\Delta)z)}{2\sqrt{z}(1-z)}\sigma_{1}\nn\\&&-\frac{iL}{2\sqrt{z}}\sigma_{2}-\frac{iL(-1+2Ln+2p-\Delta)}{2\sqrt{1-z}}+\frac{L(-i+2Lq\alpha)}{2\sqrt{1-z}}\sigma_{3}\Big)\begin{pmatrix}b_{n,p}(z)\\c_{n,p}(z)\end{pmatrix}\,.
\eea
In the above $\{\sigma_{i}\}$ are Pauli matrices. The corresponding equation of motion for the fermionic variables $(b_{n,p}(z), c_{n,p}(z))$ are given as
\bea\label{eq:b,c.AdS2.1}
&&\Big(-2iL\sqrt{z}\,\sigma_{2}\,\p_{z}+\frac{L(1-2p(1-z)+(-1+2Ln-\Delta)z)}{2\sqrt{z}(1-z)}\sigma_{1}\nn\\
&&-\frac{iL}{2\sqrt{z}}\sigma_{2}-\frac{iL(-1+2Ln+2p-\Delta)}{2\sqrt{1-z}}+\frac{L(-i+2Lq\alpha)}{2\sqrt{1-z}}\sigma_{3}\Big)\begin{pmatrix}b_{n,p}(z)\\c_{n,p}(z)\end{pmatrix}=0\,.
\eea
These equations
 provides two first order coupled differential equation in the variable $(b_{n,p}(z), c_{n,p}(z))$. Solving for $b_{n,p}(z)$ in terms of $c_{n,p}(z)$ and its derivative, we obtained
\be\label{eq:b.AdS2.1}
b_{n,p}(z) = -\frac{i ((2 p (-1 + z) + (2 L n - \Delta) z) c_{n,p}(z) + 4 (-1 + z) z\, \p_{z}c_{n,p}(z))}{
  \sqrt{z(1 - z) }\, (2 p - \Delta + 2 L (n + i q \a))}\,,
\ee
and substituting this back into the first order derivative equation for $b_{n,p}(z)$, we obtain the second order differential equation involving field $c_{n,p}(z)$ only which is
\bea\label{eq:c.AdS2.1}
&&2z\sqrt{(1-z)}\,\p^{2}_{z}\,c_{n,p}(z)+2\frac{(2-3z)}{\sqrt{1-z}}\,\p_{z}c_{n,p}(z)+\frac{1}{8z\sqrt{(1-z)^{3}}}\Big(-4p^{2}(1-z)+z(2-\Delta)\Delta\nn\\
&&+4Lz(n(-1+\Delta)+iq\a(1-z))-4L^{2}z(n^{2}+q^{2}\a^{2}(1-z))\Big)c_{n,p}(z)\=0\,.
\eea
Comparing the above equation with the bosonic equation \eqref{eq:f.AdS2.1}, we see that two are identical and therefore admit the same solution. This is not surprising, in fact it follows from supersymmetry that $\phi$ and $Q\phi$ should obey the same equation of motion. However as we will show in section \ref{sec.BdyConds.1}  that normalizable boundary condition for $\phi$ and 
fields $(B, C)$ imply different behaviour at the boundary $z\rightarrow 1$. 

\subsection{Q-exact  deformed action}\label{Sec:Q-exact.1}

Application of the method of localization requires a positive definite $Q$-exact action. 
The action given in 
\eqref{eq:susyS.1} although $Q$-exact is not positive definite. One can instead consider adding a $Q$-exact deformations which gives positive definite contribution to the Lagrangian. For this we choose $V$ to be
\be\label{eq:Q-exactV.1}
V=\int d^{3}x\,\sqrt{g}\,\,\frac{1}{2\cosh r}\Big[\psi.{(Q\psi)^{*}}+\wt\psi.{(Q\wt\psi)^{*}}\Big]\,.
\ee
In the above ${}^{*}$ is an ordinary complex conjugation. The bosonic part of the above $Q$-exact deformation is
\be \label{qexactb}
QV|_{\text{bosonic}}=\int d^{3}x\,\sqrt{g}\,\,\Big[-\bar FF+g^{\m\n}D_{\m}\bar\phi D_{\n}\phi+\frac{i}{\cosh r}\varepsilon^{\m\r\a}K_{\a}D_{\m}\phi D_{\r}\bar\phi-q^{2}\sigma^{2}\bar\phi\phi\Big]\,.
\ee
In writing the above we have used the reality condition 
\be
F^{*}=-\bar F,\quad \phi^{*}=\bar\phi\,.
\ee
By construction the bosonic part of the $QV$ deformation is positive definite (note that $\sigma$ is purely imaginary). The fermionic part of the $Q$-exact deformation is
\bea \label{qexactf}
QV|_{\text{fermionic}}=\int d^{3}x\,\sqrt{g}&&\Big[\wt\psi\slashed D\psi+2iV^{\m}(\wt\psi\g_{\m}\psi)+iq\sigma(\wt\psi\psi)-\frac{i}{\cosh r}(V\cdot K)(\wt\psi\psi)\nn\\
&&-\frac{i}{2\cosh r}\nabla_{\m}[\varepsilon^{\m\n\rho}K_{\n}(\wt\psi\g_{\r}\psi)]-\frac{1}{2}\nabla_{\m}(\wt\psi\g^{\m}\psi)\Big]\,.
\eea

Now we have two different actions. The standard action given in (\ref{eq:susyS.1}) and 
the $Q$-exact deformed action in (\ref{eq:Q-exactV.1}).  Under the principles of 
localization, if the fields obey supersymmetric boundary conditions, then the one-loop
determinants from either of the actions should yield the same result. 
To study the effect of boundary conditions in detail we will evaluate the one-loop 
determinant of both the actions using the Green's function method. 

\noindent
{\bf Kaluza-Klein decomposition:}
As in the previous section we express the bosonic part of the $QV$ action in terms of Fourier modes \eqref{eq:FourierDecmp.1} to obtain
\bea
QV|_{\text{bosonic}}&=&\int\,dz\Big[2z\sqrt{1-z}\,f_{n,p}'(z)\bar f_{n,p}'(z)\nn\\
&&-\frac{1}{2\sqrt{1-z}}\Big(2p(z-1)+(2Ln-\Delta)z\Big)\frac{d}{dz}(f_{n,p}(z)\bar f_{n,p}(z))\nn\\
&&+\frac{1}{8z(1-z)^{3/2}}\Big(4p^{2}(1-z)-4Ln\Delta z+\Delta^{2}z+4L^{2}z(n^{2}+q^{2}(1-z)\a^{2})\Big)f_{n,p}(z)\bar f_{n,p}(z)\Big]\,.\nn\\
\eea
From this we obtain the equation of motion for $f_{n,p}(z)$ which is given as
\bea\label{eq:f.AdS2.2}
&&2z\sqrt{(1-z)}\,\p^{2}_{z}\,f_{n,p}(z)+\frac{(2-3z)}{\sqrt{1-z}}\,\p_{z}f_{n,p}(z)+\frac{1}{8z\sqrt{(1-z)^{3}}}\Big(4p(p+z)(-1+z)-z\Delta(-4+\Delta+2z)\nn\\
&&+4Lnz(-2+\Delta+z)-4L^{2}z(n^{2}+q^{2}\a^{2}(1-z))\Big)f_{n,p}(z)\=0\,.
\eea
Similarly, we express the fermionic part of the $QV$ action in terms of Fourier modes and after doing some integration by parts we obtain 
\bea
QV|_{\text{fermionic}}&=&\int^{1}_{0} dz\,\frac{1}{2}\begin{pmatrix}\tilde b_{n,p}(z)&\tilde c_{n,p}(z)\end{pmatrix}\Big(2iL\sqrt{z}\,\sigma_{2}\,\p_{z}-\frac{L(1-2p(1-z)+(-1+2Ln-\Delta)z)}{2\sqrt{z}(1-z)}\sigma_{1}\nn\\&&+\frac{iL}{2\sqrt{z}}\sigma_{2}+\frac{iL(2Ln+2p-\Delta)}{2\sqrt{1-z}}-\frac{L^{2}q\alpha}{2\sqrt{1-z}}\sigma_{3}\Big)\begin{pmatrix}b_{n,p}(z)\\c_{n,p}(z)\end{pmatrix}\,.
\eea
Following the analysis in the previous section we obtain the equation of motion of $c_{n,p}(z)$ 
\bea\label{eq:c.AdS2.2}
&&2z\sqrt{(1-z)}\,\p^{2}_{z}\,c_{n,p}(z)+\frac{(2-3z)}{\sqrt{1-z}}\,\p_{z}c_{n,p}(z)+\frac{1}{8z\sqrt{(1-z)^{3}}}\Big(4p(p+z)(-1+z)-z\Delta(-4+\Delta+2z)\nn\\
&&+4Lnz(-2+\Delta+z)-4L^{2}z(n^{2}+q^{2}\a^{2}(1-z))\Big)c_{n,p}(z)\=0\,,
\eea
with the corresponding relation between $b_{n,p}(z)$ and $c_{n,p}(z)$ which is given as
\be\label{eq:b.AdS2.2}
b_{n,p}(z) = -\frac{i ((2 p (-1 + z) + (2 L n - \Delta) z) c_{n,p}(z) + 4 (-1 + z) z\, \p_{z}c_{n,p}(z))}{
  \sqrt{z(1 - z) }\, (2 p - \Delta + 2 L (n + i q \a))}\,.
\ee

\section{Boundary conditions}\label{sec.BdyConds.1}

In this section we discuss two boundary conditions that can be chosen for the
fields in the chiral multiplet. 
The consistent choice of boundary conditions  to perform the path integral 
is the normalizable boundary conditions. 
We will show that this choice of boundary conditions is not always consistent
with supersymmetric boundary conditions.

\subsection{Normalizable boundary conditions} \label{normbcond}

The normalizable boundary conditions on 
the bosonic fields $f_{n,p}(z)$ and $\bar f_{n,p}(z)$ as $z\rightarrow 1$ are
\be\label{bos.bdy.conditions}
(1-z)^{-1/4}f_{n,p}(z)\rightarrow 0,\quad (1-z)^{-1/4}\bar f_{n,p}(z)\rightarrow 0\,.
\ee
Following \eqref{twistVar.1}, we find that the normalizable boundary condition on $\psi$ and $\wt\psi$ as $z\rightarrow 1$ imposes the  following boundary conditions on twisted variables $b_{n,p},\,\wt b_{n,p}$ and $c_{n,p},\,\wt c_{n,p}$\footnote{Note that these boundary conditions follow from the standard normalizable fall off behaviour near the boundary of $AdS_2$ of the boson $\phi$ and the fermion $\psi$ in the 
chiral multiplet which are given by 
\begin{equation}
\lim_{r\rightarrow \infty}  e^{r/2}  \phi \sim  0 ,  \qquad
\lim_{r\rightarrow \infty}  e^{r/2} \psi \sim 0\,.
\end{equation}
}
\be\label{bcfermion}
b_{n,p}(z)\rightarrow 0,\quad \wt b_{n,p}(z)\rightarrow 0,\quad c_{n,p}(z)\rightarrow 0,\quad \wt c_{n,p}(z)\rightarrow 0\,.
\ee
These are also the boundary conditions that ensure that 
 boundary terms that occur on integration by parts vanish. 
 Thus the path integral is well defined with these boundary conditions. 
 It can be seen that these boundary conditions together with the smoothness conditions for the fields in the chiral multiplet near the origin of $AdS_2$ ensure that the boundary term given in 
(\ref{boundterm}),
that arises on writing the standard action as a $Q$-exact action, vanishes. 
Furthermore, these  boundary conditions also need to be imposed 
on the $Q$-exact deformed action given in (\ref{qexactb}), (\ref{qexactf}) 
 to define the path integral.

The one loop determinants for the operators in the standard action as well as the 
Q-exact action therefore should be evaluated on the space of solutions 
satisfying the normalizable boundary conditions. 
Without knowing the explicit form of the solutions of the differential equations
(\ref{eq:c.AdS2.1}), (\ref{eq:c.AdS2.1}), (\ref{eq:f.AdS2.2}), (\ref{eq:c.AdS2.2}) 
by studying the asymptotic behaviour 
of the solutions we can obtain those that satisfy normalizable 
boundary conditions. 
Let us discuss this first  for   the solutions of the equations of motion of the 
standard action. 
From  studying the roots of the indicial equation of (\ref{eq:c.AdS2.1}) at 
$z=1$ we see that the two  behaviours for $f_{n, p}$ are given by 
\begin{eqnarray} \label{bcfnp}
f_{n, p}^+  &\sim&  ( 1- z)^{\frac{1}{4}( 2Ln -\Delta + 2) } + \cdots  , \\ \nonumber
f_{n, p}^-  &\sim&  ( 1- z)^{\frac{1}{4}( -2Ln +\Delta ) }+ \cdots \,.
\end{eqnarray}
Therefore for  the solutions which are normalizable, that is which satisfies 
the conditions (\ref{bos.bdy.conditions}) is given by 
\begin{equation}\label{norm-f}
f_{n, p}|_{\rm normalizable} = 
\begin{cases} f_{n, p}^+ , \qquad  \hbox{for}\; \; n> \frac{\Delta -1}{2L} \\
f_{n, p}^- , \qquad  \hbox{for}\; \; n< \frac{\Delta -1}{2L}\,.
\end{cases}
\end{equation}
The asymptotic behaviour of 
$c_{n, p}$  is the same as that of $f_{n, p}$ since its equation given in    (\ref{eq:c.AdS2.1}) is identical 
to that of $f_{n, p}$.  
\begin{eqnarray}\label{assce}
c_{n, p}^+ &\sim& ( 1- z)^{\frac{1}{4}( 2Ln -\Delta + 2) }+ \cdots , \\ \nonumber
c_{n, p}^+ &\sim& ( 1- z)^{\frac{1}{4}( -2Ln -\Delta ) }+ \cdots\,.
\end{eqnarray}
Therefore, solutions which satisfy the normalizable 
boundary conditions in (\ref{bcfermion})  are given by 
\begin{equation} \label{norm-c}
c_{n, p}|_{\rm normalizable} =
\begin{cases} c_{n, p}^+ , \qquad  \hbox{for}\; \; n> \frac{\Delta -2}{2L} \\
c_{n, p}^- , \qquad  \hbox{for}\; \; n< \frac{\Delta}{2L}\,.
\end{cases} 
\end{equation}
Note that both $c_{n, p}^+$ as well as $c_{n, p}^-$ are admissible in the interval
$\frac{\Delta -2}{2L} <n < \frac{\Delta}{2L}$. 
Finally using  (\ref{assce})  and (\ref{eq:b.AdS2.1}) we   obtain the asymptotic 
behaviour
\begin{eqnarray}\label{bccb}
b_{n, p}^+& \sim& (1-z) ^{\frac{1}{4} ( 2Ln - \Delta) } + \cdots, \\ \nonumber
b_{n, p}^- &\sim&  (1-z) ^{\frac{1}{4} ( -2Ln + \Delta +2) } + \cdots\,. \\ \nonumber
\end{eqnarray}
Therefore the solutions which satisfies the boundary conditions (\ref{bcfermion}) are given by
\begin{eqnarray}\label{norm-b}
b_{n, p}|_{\rm normalizable} = 
\begin{cases} b_{n, p}^+ , \qquad  \hbox{for}\; \; n> \frac{\Delta}{2L} \\
b_{n, p}^- , \qquad  \hbox{for}\; \; n< \frac{\Delta+2}{2L}\,.
\end{cases} 
\end{eqnarray}
Again  both $b_{n, p}^+$ as well as $b_{n, p}^-$  are admissible in the interval
$\frac{\Delta }{2L} <n < \frac{\Delta+2}{2L}$. 
Combining  (\ref{norm-f}), (\ref{norm-b}) and (\ref{norm-c}) we see that the solutions for the 
system 
$\{ f,  (b, c) \} $ which satisfies normalizable boundary conditions are 
\begin{eqnarray}\label{finalnorm}
 \{ f_{n, p} , ( b_{n, p}, c_{n, p}) \}|_{\rm normalizable} 
=  \begin{cases}
  \{ f^+, ( b_{n, p}^+, c_{n, p}^+  ) \}  \;\; \hbox{for}\; \; n> \frac{\Delta}{2L} \\
\{ f^+, ( b_{n, p}^-, c_{n, p}^-  ) \}   \;\;\hbox{for}\; \;  \frac{\Delta-1}{2L} <n<\frac{\Delta}{2L} \\
\{ f^-, ( b_{n, p}^-, c_{n, p}^-  ) \}  \;\; \hbox{for}\;\; n <  \frac{\Delta-1}{2L} \,.
\end{cases}
\end{eqnarray}
To arrive at this conclusion it is important to use the fact that 
the $(b, c)$ system must  satisfy (\ref{eq:b.AdS2.1}) and therefore, both are either
the `$+$' modes or  both   `$-$' modes. 

\subsection{Supersymmetric boundary condtions}

The normalizable boundary conditions discussed in the previous
section  are not consistent with supersymmetry. 
Following supersymmetry one can also impose boundary conditions which closes under supersymmetry transformations. Using the normalizable boundary condition on the bosonic field \eqref{bos.bdy.conditions} and the following supersymmetry transformations
\bea
&&Qf_{n,p}(r)=c_{n,p}(r),\quad Q\bar f_{n,p}(r)=\wt c_{n,p}(r)\,,\nn\\
&&Qb_{n,p}=\frac{1}{4L\sinh r}\left[(2Ln+4p-\Delta+(-2Ln+\Delta)\cosh2r)f_{n,p}(r)-2\sinh 2r(L F_{n,p}-\p_{r} f_{n,p}(r))\right]\,,\nn\\
&&Q\wt b_{n,p}=-\frac{1}{4L\sinh r}\left[(-2Ln-4p+\Delta+(2Ln-\Delta)\cosh2r)\bar f_{n,p}(r)+2\sinh 2r(L\bar F_{n,p}-\p_{r}\bar f_{n,p}(r))\right]\,,\nn\\
\eea
one can impose the following supersymmetric boundary condition on $b,\,\tilde b$ and $c,\,\tilde c$ as $z\rightarrow 1$ is
\be\label{susybc1}
(1-z)^{1/4}b_{n,p}(z)\rightarrow 0,\quad (1-z)^{1/4}\wt b_{n,p}(z)\rightarrow 0,\quad (1-z)^{-1/4}c_{n,p}(z)\rightarrow 0,\quad (1-z)^{-1/4}\tilde c_{n,p}(z)\rightarrow 0\,.
\ee
We see from the above boundary conditions that the supersymmetry transformations allows the modes for $b_{n,p}(z)$ and $\wt b_{n,p}(z)$ to diverge as $z\rightarrow 1$.
Note that the boundary term  (\ref{boundterm}) that arises on writing the standard action 
as a $Q$-exact term also vanishes using supersymmetric boundary conditions. 
The way to see this is to observe that  under the conditions (\ref{susybc1}) we have
\begin{equation}
\lim_{r\rightarrow \infty} e^{\frac{r}{2} }\tilde\epsilon \tilde\psi  = 0 , 
\quad \lim_{r\rightarrow \infty }  (\epsilon \gamma^\mu \psi)  \sim e^{\frac{r}{2} }\,.
\end{equation}
These equations can be obtained by using (\ref{susybc1}) along with (\ref{twistVar.1}) and
(\ref{twist2}). 
With this behaviour the first term in boundary term in 
 (\ref{boundterm}) vanishes, while the rest of the terms vanish on the account of the 
 normalizable boundary condition on the bosonic field $f$. 
Using the asymptotic behaviours in (\ref{bcfnp}), (\ref{bccb}) and (\ref{assce})  and 
going through the same analysis as in the previous section we obtain the following admissible 
solutions for  the  fields. 
\begin{eqnarray}
f_{n, p}|_{\rm susy} = 
\begin{cases} f_{n, p}^+ , \qquad  \hbox{for}\; \; n> \frac{\Delta -1}{2L} \\
f_{n, p}^- , \qquad  \hbox{for}\; \; n< \frac{\Delta -1}{2L}
\end{cases} \\ \nonumber
c_{n, p}|_{\rm susy} = 
\begin{cases} c_{n, p}^+ , \qquad  \hbox{for}\; \; n> \frac{\Delta -1}{2L} \\
c_{n, p}^- , \qquad  \hbox{for}\; \; n< \frac{\Delta -1}{2L}
\end{cases} \\ \nonumber
b_{n, p}|_{\rm susy} = 
\begin{cases} b_{n, p}^+ , \qquad  \hbox{for}\; \; n> \frac{\Delta -1}{2L} \\
b_{n, p}^- , \qquad  \hbox{for}\; \; n< \frac{\Delta +3}{2L}\,.
\end{cases} 
\end{eqnarray}
Now using these solutions, the combined system satisfies 
supersymmetric boundary conditions when
\begin{equation} \label{finalsusy}
\{ f_{n, p} , ( b_{n, p}, c_{n, p}) \}|_{\rm susy} 
=  \begin{cases}
  \{ f^+, ( b_{n, p}^+, c_{n, p}^+  ) \}  \;\; \hbox{for}\; \; n> \frac{\Delta-1}{2L} \\
\{ f^-, ( b_{n, p}^-, c_{n, p}^-  ) \}   \;\;\hbox{for}\; \;  n<\frac{\Delta-1}{2L} \,.
\end{cases}
\end{equation}

Comparing (\ref{finalnorm}) and (\ref{finalsusy})  we see that fields  which 
satisfy  normalizable boundary conditions
also satisfy  supersymmetric boundary conditions unless there exists
an integer $n$ in the open  interval
\begin{equation}\label{domain}
 D: \quad \Big( \frac{\Delta -1}{2L}, \frac{\Delta}{2L} \Big) \,.
\end{equation}
It is only if there is exits integers in the interval $D$, the Kaluza-Klein modes 
which are normalizable do not obey supersymmetric boundary conditions. 
Now by the principle of localization we expect that 
if there exist no integer $n$  in the interval $D$, the one loop determinant 
 of  the actions (\ref{eq:susyS.1}) and (\ref{qexactb}), (\ref{qexactf}) 
  evaluated on solutions obeying normalizable boundary conditions 
 to agree.  This is because in this case they will also obey supersymmetric boundary 
 conditions.  Now if there exist an integer $n$  in the interval $D$  we expect the 
 final one loop determinants of the standard action and the $Q$-exact deformed 
 action to no longer agree. 
 We will demonstrate this explicitly in the next section.

\section{ One loop determinants from the Green's functions} \label{1-loop}

Note that the action in (\ref{eq:susyS.1}) is quadratic in the fields of the chiral multiplet. 
Therefore, given the  background vector multiplet  in (\ref{backvec}) we can in principle obtain
its one loop determinant. 
The general rules of quantum field theory dictates that we use normalizable boundary 
conditions for the fields in the path integral. 
However the method of localization will require the boundary conditions 
also to  be consistent with supersymmetry. 
We have shown in the previous section that these two boundary conditions 
are not consistent if there exists an integer $n$  in the domain (\ref{domain}). 
 The action given in (\ref{eq:susyS.1}) is a simple example of a situation in which one can evaluate the one loop determinant using 
the method of localization as well as directly by using the Green's function approach. 
The one loop determinant calculation using Green's function approach relies on the boundary conditions being normalizable 
while the method of localization relies on the boundary conditions 
being supersymmetric. Thus if these boundary conditions are not 
consistent the answer for the one loop determinant will in general be different. 
It is this phenomenon we wish to make explicit.  
We will do this by  carrying out the following steps :
\begin{enumerate}
\item  Evaluate the one loop determinant of the action (\ref{eq:susyS.1}) in the 
vector multiplet background (\ref{backvec})  by developing the Green's function approach. 
This relies on normalizable boundary conditons. 
\item Evaluate the one loop determinant of the $Q$-exact action (\ref{qexactb}), (\ref{qexactf}) in the 
vector multiplet background (\ref{backvec})  again using the Green's function approach. 
\item Comparing the results we will see that  one loop determinants of the 
standard action and that of the $Q$-exact action 
  differ only when there exists
an integer $n$  in the domain (\ref{domain}). 
\item Finally  in section \ref{indexd10} we obtain  the one loop determinant 
by evaluating the index of the $D_{10}$ operator corresponding to the 
localizing  action given in (\ref{eq:Q-exactV.1}).  The method relies on 
using boundary conditions which are supersymmetric. 
We observe that  this one loop determinant differs from that 
 obtained from the Green's function  for the Q-exact action 
 when there 
exists an integer $n$ in the domain (\ref{domain}). 
\end{enumerate}

\subsection{One loop determinant of the standard  action}\label{1loopst}

To evaluate the one loop determinant of the action in 
(\ref{eq:susyS.1}) we first obtain the Green's function  for the bosonic 
and fermionic operators  present in this action.
Note that the action depends on the parameter $\alpha$ 
(\ref{backvec}) which parametrises the expectation value of the 
scalar of the vector multiplet. 
Therefore, the one loop determinant  will depend on $\alpha$. 
Now  by the standard rules  which relate the one loop determinant 
to the Green's function we have the equation
\begin{equation}\label{greendet}
\frac{\delta}{\delta \alpha} \ln Z_{\rm 1-loop} (\a)
= {\rm Tr} [ G_F \frac{\delta}{\delta \alpha} {\cal D}_F(\alpha) ]
- \frac{1}{2} {\rm Tr} [ G_B \frac{\delta}{\delta \alpha} {\cal D}_B(\alpha) ]\,.
\end{equation}
Here $G_B$ is the Green's function of the bosonic operator ${\cal D}_B$, while 
$G_{F}$ is the Green's function of the fermionic operator ${\cal D}_F$ which 
occurs in the action (\ref{eq:susyS.1}). 
Once the Green's functions are known, we can  use  (\ref{greendet}) and integrate 
with respect to $\alpha$ and obtain the one loop determinant up to a constant 
independent of $\alpha$. 
In fact for specific values of $\Delta, L$ we will show that that this constant is trivial
by directly evaluating the one loop determinant with $\alpha=0$ using the eigen function
method. 

The Green's functions for  bosonic and fermionic operators of the action (\ref{eq:susyS.1})
are constructed by solving the differential equations satisfied by the corresponding 
fields after the Kaluza-Klein reduction. For the bosonic operator, this is 
given in (\ref{eq:f.AdS2.1}) and for the fermionic operator it is given in (\ref{eq:b,c.AdS2.1}). 
Note that
 differential equation for the boson  $f_{n,p}(z)$, \eqref{eq:f.AdS2.1}, and  fermion $c_{n,p}(z)$,  \eqref{eq:c.AdS2.1}, 
 are same.  
 Since both the equations are same, we just need to find the solution for one of the differential equation to obtain the Green's function. 
  However we need to make sure that the corresponding solution for $b_{n,p}(z)$ obtained by using \eqref{eq:b.AdS2.1} is smooth in the interior and satisfy the right boundary conditions. 
The solution which is smooth near $z=0$ for $p>0$ is
\be
S_{1+}(z)=(1 - z)^{\frac{1}{4}( - 2 L n + \Delta)} z^{p/
  2}\,{}_2F_{1}[ \frac{1}{4}(2 - 2 L n + 2 p + \Delta + 2 i L q \alpha), \frac{1}{4} (2 p + \Delta - 2 L (n + i q \alpha)), 1 + p, z]\,,
\ee
which we see by noticing that its $z\rightarrow 0$ behaviour goes like
\be
S_{1+}(z)\rightarrow z^{p/2}+\mathcal O(z^{p/2+1})\,.
\ee
The solution which is smooth near $z=0$ for $p<0$ 
\be
S_{1-}(z)=(1 - z)^{\frac{1}{4}( - 2 L n + \Delta)} z^{-p/
  2} \, {}_2F_{1}[
     \frac{1}{4} (2 - 2 L n - 2 p + \Delta + 2 i L q \alpha), \frac{1}{4} (-2 p + \Delta - 2 L (n + i q \alpha), 1 - p, z] \,,
\ee
which goes like, for $z\rightarrow 0$
\be
S_{1-}(z)\rightarrow z^{-p/2}+\mathcal O(z^{-p/2+1})\,.
\ee
For $p=0$ both the solutions are identical and we take this to be $S_{1-}(z)$. In this case the second independent solution is singular which goes logarithmic near $z=0$.\\
Now to get the behaviour of the solution at $z=1$, we start with the differential equation \eqref{eq:f.AdS2.1} and substitute $z\rightarrow 1-y$, then we get
\bea
&&\Big(2 \Delta (1 - y) - \Delta^2 (1 - y) - 4 p^2 y - 4 L (1 - y) (n - n \Delta - i q y \alpha) - 4 L^2 (1 - y) (n^2 + q^2 y \alpha^2)\Big) f_{n,p}(y)\nn\\ 
&&+ 8 (1 - y) y \Big((1 - 3 y) \p_{y}f_{n,p}(y) + 2 (1 - y) y \,\p^{2}_{y}f_{n,p}(y)\Big)=0\,.\nn\\
\eea
In this case we again have two solutions. The solution which satisfy normalizable boundary condition for $n>\frac{\Delta-1}{2L}$ is
\bea
S_{2+}(z)&=&(1 - z)^{\frac{1}{4}( 2+ 2 L n - \Delta)} z^{-p/2}\times\nn\\&&{}_2F_{1}[ \frac{1}{4}(2 + 2 L n - 2 p - \Delta - 2 i L q \alpha), \frac{1}{4} (4-2 p - \Delta + 2 L (n + i q \alpha)), \frac{3}{2}+Ln-\frac{\Delta}{2}, 1-z]\,,\nn\\
\eea
and the one which satisfies normalizable boundary condition for $n<\frac{\Delta-1}{2L}$ is
\bea
S_{2-}(z)&=&(1 - z)^{\frac{1}{4}(-2 L n + \Delta)} z^{-p/2}\times\nn\\&&{}_2F_{1}[ \frac{1}{4}(2 - 2 L n - 2 p + \Delta + 2 i L q \alpha), \frac{1}{4} (-2 p + \Delta - 2 L (n + i q \alpha)), \frac{1}{2}(1-2Ln+\Delta), 1-z]\,.\nn\\
\eea
For $n=\frac{\Delta-1}{2L}$, both the solutions coincide. However in this case
\be
S_{2+}(z)=S_{2-}(z)\sim (1-z)^{1/4}+...\,.
\ee
So for this value of $n$, $f_{n,p}(z)$ is at the border of normalizability. The second independent solution involves a logarithmic divergence.\\

Now we look for the solution for the fermionic equations. Since the equation for $c(z)$ is same as that of $f(z)$, we will , therefore, have the same solutions. 
As discussed in section \ref{normbcond}  the normalizable boundary conditions for 
$b(z)$ and $c(z)$  are given by 
\be
b(z)\rightarrow 0,\quad c(z)\rightarrow 0\,,\quad \text{for} \,\,z\rightarrow 1\,.
\ee
Therefore, $S_{2+}(z)$ is valid solution for $c(z)$ for $n>\frac{\Delta-2}{2L}$ and $S_{2-}(z)$ is valid solution for $n<\frac{\Delta}{2L}$. Thus we see that in the range $\frac{\Delta-2}{2L}<n<\frac{\Delta}{2L}$, both solutions $S_{2+}(z)$ and $S_{2-}(z)$ are admissible solutions for $c(z)$. Now we will substitute the expression for $c(z)$ in \eqref{eq:b.AdS2.1} to obtain $b(z)$ and find its asymptotic behaviour. One gets
\bea
&&c_{n,p}(z)=S_{2+}(z),\qquad b_{n,p}(z)\sim (1-z)^{\frac{1}{4}(2Ln-\Delta)}+...\nn\\
&&c_{n,p}(z)=S_{2-}(z),\qquad b_{n,p}(z)\sim (1-z)^{\frac{1}{4}(-2Ln+\Delta +2)}+...\,.
\eea
So for $b(z)$, $S_{2+}(z)$ is valid solution for $n>\frac{\Delta}{2L}$ and $S_{2-}(z)$ is valid solution for $n<\frac{\Delta +2}{2L}$. So for $(b,c)$ system, we have valid solution $S_{2+}(z)$ for $n>\frac{\Delta}{2L}$ and $S_{2-}(z)$ is valid solution for $n<\frac{\Delta}{2L}$. 
So we see that for $n>\frac{\Delta}{2L}$ and $n<\frac{\Delta-1}{2L}$ both $f(z)$ and $(b(z),c(z))$ have the same admissible solutions. However there is a mismatch of the solution in the range $\frac{\Delta-1}{2L}<n<\frac{\Delta}{2L}$. In this range the valid solution for $f(z)$ is $S_{2+}(z)$ whereas for $(b(z),c(z))$ system, the valid solution is $S_{2-}(z)$.  
This conclusion was reached earlier  in section \ref{normbcond}
just by the analysis of 
the asymptotic properties of the solutions.   Here we have the explicit solutions of the 
differential equations \eqref{eq:f.AdS2.1} and  (\ref{eq:b,c.AdS2.1}) and the necessary 
choices of functions so as  to satisfy the boundary conditions at $z=0$ as well 
at $z=1$ . Using these solutions we can construct the Green's functions corresponding to these
differential equations. \\
We still need to analyse the boundary behaviour of the mode at $n=\frac{\Delta}{2L}$ (if it is an integer). At $n=\frac{\Delta}{2L}$ we have a similar feature as noticed in the bosonic case at $n=\frac{\Delta-1}{2L}$. If we consider $c_{\frac{\Delta}{2L},p}(z)=S_{2+}(z)$ which goes like $\sqrt{1-z}$ as $z\rightarrow 1$, the corresponding mode for $b_{\frac{\Delta}{2L},p}(z)$ goes like $\mathcal O(1)$ and is at the border of normalizability. On the other hand for $c_{\frac{\Delta}{2L},p}(z)=S_{2-}(z)$ which goes like $\mathcal O(1)$ as $z\rightarrow 1$, the corresponding mode for $b_{\frac{\Delta}{2L},p}(z)$ goes like $\sqrt{1-z}$ and is admissible. Thus for the $(b,c)$ system none of the modes at $n=\frac{\Delta}{2L}$ satisfy strict normalizable boundary conditions.

\noindent
{\bf  Construction of the Green's function: $n>\frac{\Delta}{2L}$ or $n<\frac{\Delta-1}{2L}$}\\
In this case both $f(z)$ and $(b(z),c(z))$ have same valid solution as $z\rightarrow 1$, lets call it $S_{2}(z)$. Also let us 
 assume that the smooth solution near $z\rightarrow 0$ is $S_{1}(z)$, it will be $S_{1\pm}(z)$ depending on whether $p$ is positive or negative or zero in which case both, $S_{1+}(z)$ and $S_{1-}(z)$, are equal. 
Let us first construct the  Green's function for bosonic  field $f_{n, p}$.   
Given the solutions of \eqref{eq:f.AdS2.1}, the Green's function can be written as
\be\label{eq:G.Bos}
G_{b}(z,z')=c_{b}\Big[\Theta(z'-z)\,S_{1}(z)S_{2}(z')+\Theta(z-z')S_{1}(z')S_{2}(z)\Big]\,.
\ee
$c_{b}$ is some constant.\\
This satisfies the continuity condition 
\be\label{continuity}
\lim_{\epsilon\rightarrow 0}G(z'-\epsilon,z')=\lim_{\epsilon\rightarrow 0}G(z'+\epsilon,z')\,.
\ee
The discontinuity condition for the first derivative of Green's function implies that
\be \label{discont}
S_{1}(z')\p_{z}S_{2}(z)|_{z=z'}-S_{2}(z')\p_{z}S_{1}(z)|_{z=z'}=\frac{1}{c_{b}\,a(z')}\,,
\ee
where
\be
a(z)=-2z\sqrt{1-z}\,.
\ee
To fix the constant $c_b$ we consider 
 the Wronskian which is defined as 
\be \label{W1}
W(z)=\p_{z}S_{2}(z)\,S_{1}(z)-\p_{z}S_{1}(z)\,S_{2}(z)\,.
\ee
The Wr\"{o}nskian satisfies
\be\label{W2}
\p_{z}W(z)-\frac{3z-2}{2z(z-1)}W(z)=0\,.
\ee 
The solution is given by
\be\label{Sol;W}
W(z)=\frac{c_{1}}{z\sqrt{1-z}}\,.
\ee
Thus comparing with the condition  (\ref{discont})   the constants 
constants $c_1$ and $c_b$  are related by 
\be \label{defcb}
c_{b}=\frac{z\sqrt{1-z}}{c_{1}\,a(z)}=-\frac{1}{2c_{1}}\,.
\ee

The constant $c_{1}$ can be determined by evaluating $W(z)$ at some value of $z$. Its value depends on integers $(n,p)$. 
We  now 
 will determine the constant $c_{1}$ for various cases.
When $S_{1}(z)=S_{1+}(z)$ and $S_{2}(z)=S_{2+}(z)$, then evaluating $W(z)$ near $z\rightarrow 0$, we obtain
\be
\lim_{z\rightarrow 0}W(z)=-p\frac{\Gamma(p)\Gamma(\frac{3}{2}+Ln-\frac{\Delta}{2})}{z\,\Gamma\Big[\frac{1}{4}(2+2p+2L(n-iq\alpha)-\Delta)\Big]\Gamma\Big[\frac{1}{4}(4+2p+2L(n+iq\alpha)-\Delta)\Big]}\,,
\ee
and comparing with \eqref{Sol;W} we obtain the constant
\be\label{case1}
c_{1++}=-p\frac{\Gamma(p)\Gamma(\frac{3}{2}+Ln-\frac{\Delta}{2})}{\Gamma\Big[\frac{1}{4}(2+2p+2L(n-iq\alpha)-\Delta)\Big]\Gamma\Big[\frac{1}{4}(4+2p+2L(n+iq\alpha)-\Delta)\Big]}\,.
\ee
 Similarly, when $S_{1}(z)=S_{1-}(z)$ and $S_{2}(z)=S_{2+}(z)$, then by evaluating $W(z)$ near $z\rightarrow 0$, we obtain
\be\label{case2}
c_{1-+}=p\frac{\Gamma(-p)\Gamma(\frac{3}{2}+Ln-\frac{\Delta}{2})}{\Gamma\Big[\frac{1}{4}(2-2p+2L(n-iq\alpha)-\Delta)\Big]\Gamma\Big[\frac{1}{4}(4-2p+2L(n+iq\alpha)-\Delta)\Big]}\,.
\ee
When $S_{1}(z)=S_{1+}(z)$ and $S_{2}(z)=S_{2-}(z)$, then by evaluating $W(z)$ near $z\rightarrow 0$, we obtain
\be\label{case3}
c_{1+-}=-p\frac{\Gamma(p)\Gamma(\frac{1}{2}(1-2Ln+\Delta)}{\Gamma\Big[\frac{1}{4}(2+2p-2L(n-iq\alpha)+\Delta)\Big]\Gamma\Big[\frac{1}{4}(2p-2L(n+iq\alpha)+\Delta)\Big]}\,.
\ee
When $S_{1}(z)=S_{1-}(z)$ and $S_{2}(z)=S_{2-}(z)$, then by evaluating $W(z)$ near $z\rightarrow 0$, we obtain
\be\label{case4}
c_{1--}=p\frac{\Gamma(-p)\Gamma(\frac{1}{2}(1-2Ln+\Delta))}{\Gamma\Big[\frac{1}{4}(2-2p-2L(n-iq\alpha)+\Delta)\Big]\Gamma\Big[\frac{1}{4}(-2p-2L(n+iq\alpha)+\Delta)\Big]}\,.
\ee
These constants will play important role in calculating the one loop determinant.
For instance for $p>0$ and  $n> \frac{\Delta}{2L}$, the Green's function in \ref{eq:G.Bos}
will involve the constant $c_b = -\frac{1}{2c_{1++}}$.  Similar statements apply for 
all the other cases in (\ref{case2}), (\ref{case3}) and (\ref{case4})\,.

Let us now  compute the Green's function for fermions. 
The  Green's  function  for the $(b,c)$ system satisfies the first order differential equation
\be\label{matrixg}
\begin{pmatrix}A&&B\\C&&D\end{pmatrix}G_{f}(z,z')=\delta(z-z')\,,
\ee
where the various elements of the matrix are given in \eqref{eq:b,c.AdS2.1}. 
The Green's function  can then be written in terms of the solutions to the 
$b, c$ system  as 
\be \label{fgreen}
G_{f}(z,z')=c_{f}\Big[\Theta(z'-z)\begin{pmatrix}b_{1}(z)b_{2}(z')&&b_{1}(z)S_{2}(z')\\b_{2}(z')S_{1}(z)&&S_{1}(z)S_{2}(z')\end{pmatrix}+\Theta(z-z')\begin{pmatrix}b_{1}(z')b_{2}(z)&&b_{2}(z)S_{1}(z')\\b_{1}(z')S_{2}(z)&&S_{1}(z')S_{2}(z)\end{pmatrix}\Big]\,.
\ee
Here $b_{1}(z)$ and $b_{2}(z)$ are determined from \eqref{eq:b.AdS2.1} by substituting $c(z)\rightarrow S_{1}(z)$ and $c(z)\rightarrow S_{2}(z)$ respectively.  
What is left in the construction of the fermionic Green's function is to determine the 
constant $c_f$.  We now relate $c_f$ to the bosonic constant $c_b$ in (\ref{defcb})\,.
 From (\ref{matrixg}) we see 
 that  the continuity 
constraint of the Green's function given in (\ref{continuity}) needs to be satisfied only by the 
diagonal elements. 
Evaluating the discontinuity  of   the Green's function in (\ref{fgreen}) 
we obtain
\begin{equation}
\lim_{\epsilon\rightarrow 0} ( G_f(z'-\epsilon, z' ) - G_f( z' + \epsilon, z') )  = 
-\frac{4c_{f}c_{1}}{\sqrt{z'}(2Ln+2p-\Delta+2iLq\alpha)}\sigma_{2}\,.
\end{equation}
To obtain this we have also used the relation obtained from the Wr\"{o}nksian 
(\ref{W1})  and (\ref{Sol;W}) which is given by 
\be
\p_{z}S_{2}(z)=\frac{c_{1}+z\sqrt{1-z}S_{2}(z)\p_{z}S_{1}(z)}{z\sqrt{1-z}\,S_{1}(z)}\,.
\ee
Integrating the 
differential equation  (\ref{matrixg})   from $z'-\epsilon$ to $z'+\epsilon$ and 
from the fact 
that the first order derivative in \eqref{eq:b,c.AdS2.1}
comes with the  coefficient  $-2iL\sqrt{z}\,\sigma_{2}$ we obtain 
\be\label{deffc}
\frac{8iLc_{f}c_{1}}{2Ln+2p-\Delta+2iLq\alpha}=1\Rightarrow c_{f}=-\frac{(2Ln+2p-\Delta+2iLq\alpha)c_{b}}{4iL}\,.
\ee
where we have used $c_{1}=-\frac{1}{2c_{b}}$.
This completes the construction of the fermionic Green's function.  It is important to 
note that for each of the cases in (\ref{case1}), (\ref{case2}), (\ref{case3}), (\ref{case4}) we use 
(\ref{deffc}) to obtain the corresponding values of $c_f$. 

We are now ready to compute the contribution to the  one loop determinant 
from  the modes in the range $n> \Delta/2L$ and $ n< (\Delta - 1)/2L$. 
If  $\mathcal D_{b}(\alpha)$ and $\mathcal D_{f}(\alpha)$ are the differential operator for the complex scalar and fermions, respectively whose determinant we are interested in to calculate and $\alpha$ is the background value of scalar field in the vector multiplet, then\footnote{Note that there is no factor of $\frac{1}{2}$ in the last term because we are considering a complex scalar.}
\bea
&&\frac{\delta}{\delta\alpha}\ln \wt Z_{1-loop}(\a)=\text{Tr}[G_{F}\frac{\delta}{\delta\alpha}\mathcal D_{f}(\alpha)]-\text{Tr}[G_{b}\frac{\delta}{\delta\alpha}\mathcal D_{b}(\alpha)]\,.
\eea
Here $\wt Z_{1-loop}(\a)$  is the contribution to one loop determinant coming from the modes in the range $n> \Delta/2L$ and $ n< (\Delta - 1)/2L$\,. Now for the specific differential operators of our interest we obtain 
\be
\frac{\delta}{\delta\alpha}\mathcal D_{f}(\alpha)=\frac{L^{2}q}{2\sqrt{1-z}}\sigma_{3},\quad \frac{\delta}{\delta\alpha}\mathcal D_{b}(\alpha)=\frac{Lq(-i+2Lq\alpha)}{2\sqrt{1-z}}\,.
\ee
Evaluating the difference in the Green's function  and performing the trace in 
position space we obtain 
\bea
\frac{\delta}{\delta\alpha}\ln \wt Z_{1-loop}(\a)&=&\int^{1}_{0} dz\frac{c_{b} L q}{4 \sqrt{(1 - z)}} \Big[2 (i - 2 L q \alpha) S_{1}(z) S_{2}(z) - i (2 p -\Delta + 2 L (n + i q \alpha)) \nn\\&&\times \Big(S_{1}(z) S_{2}(z)+ \frac{XY}{(1 - z) z (-2 p + \Delta - 2 L (n + i q \alpha))^2}\Big)\Big]\,.
\eea
Here
\bea
X=(2 p (-1 + z) + (2 L n - \Delta) z) S_{1}(z) + 4 (-1 + z) z \p_{z}S_{1}(z)\,,\nn\\
Y=(2 p (-1 + z) + (2 L n - \Delta) z) S_{2}(z) + 4 (-1 + z) z \p_{z}S_{2}(z)\,.
\eea   
After integrating by parts \footnote{Note that integrating by parts is allowed since all the 
fields  satisfy normalizable boundary conditions.}
and using the fact that $S_{1}(z)$ satisfies equation of motion, we find that the integrand is
a  total derivative and therefore the integral is given by
\be\label{eq:Bdy.AdS2.1}
\frac{\delta}{\delta\alpha}\ln\wt Z_{1-loop}=-\frac{i L q S_{2}(z) ((2 p (-1 + z) + (2 L n - \Delta) z) S_{1}(z) + 4 (-1 + z) z \p_{z}S_{1}(z))}{2 \sqrt{1 - z} (2 p - \Delta + 2 L (n + iq \alpha)) c_{1}}\Big|^{1}_{0}\,.
\ee
We now evaluate the right hand side of the above expression for various range of $(n,p)$. 
We see that the expression depends on the integration constant $c_1$ from the 
denominator.
However the numerator also proportional to $c_1$. Therefore the expression 
in independent of $c_1$ and in fact we do not need the explicit expressions
given in (\ref{case1})-(\ref{case4})  to evaluate it. 
Let us  denote the right hand side of the  expression in (\ref{eq:Bdy.AdS2.1}) 
 by $\text{BT}$. \\
${n>\frac{\Delta}{2L}}$ and ${p>0}:$ we have $S_{1}(z)=S_{1+}(z)$ and $S_{2}(z)=S_{2+}(z)$. In this case the boundary term is  given by 
\be\label{eq:bdyterm.1}
\text{BT}=\frac{2iLq}{2p-\Delta+2L(n+iq\alpha)}\,.
\ee
${n>\frac{\Delta}{2L}}$ and ${p\leq0}:$ we have $S_{1}(z)=S_{1-}(z)$ and $S_{2}(z)=S_{2+}(z)$. In this case the boundary term is  given by 
\be\label{eq:bdyterm.2}
\text{BT}=0\,.
\ee
${n<\frac{\Delta-1}{2L}}$ and ${p>0}:$ we have $S_{1}(z)=S_{1+}(z)$ and $S_{2}(z)=S_{2-}(z)$. In this case the boundary term is  given by 
\be\label{eq:bdyterm.3}
\text{BT}=0\,.
\ee
${n<\frac{\Delta-1}{2L}}$ and ${p\leq0}:$ we have $S_{1}(z)=S_{1-}(z)$ and $S_{2}(z)=S_{2-}(z)$. In this case for the boundary term is  given by 
\be\label{eq:bdyterm.4}
\text{BT}=-\frac{2iLq}{2p-\Delta+2L(n+iq\alpha)}\,.
\ee
We can integrate each of these expressions with respect to $\alpha$ to obtain the 
one loop determinant. 

\noindent 
{\bf{ Construction of the Green's function:}}  $\frac{\Delta-1}{2L}<n<\frac{\Delta}{2L}$

We still need to analyse the region when the Kaluza-Klein mode $n$ lies in the interval
 $D:\;\frac{\Delta-1}{2L}<n<\frac{\Delta}{2L}$. Of course if there is no integer $n$  in the interval $( \frac{\Delta-1}{2L}, \frac{\Delta}{2 L} )$
then the contributions to the one loop determinant  obtained by integrating 
(\ref{eq:bdyterm.1}), (\ref{eq:bdyterm.2}), (\ref{eq:bdyterm.3}), (\ref{eq:bdyterm.4}) 
 is the complete answer.
 For example  consider the situation with $\Delta=3/2$ and $L =1$ then this range is $(1/4, 3/4)$
 and  there exists no   integer  $n$ in this range. 
 In general if $\Delta>1$ and $L>1/2$ we see that there are no integers in the domain $D$. 
However, consider the case with $\Delta = 1/2$. The theory is conformal at this point, 
the domain $D$ becomes 
 $(-\frac{1}{4 L}, +\frac{1}{4L})$ and  the integer $n=0$ is necessarily lies   inside this range. 
 
 Let us evaluate the contribution of the Kaluza-Klein modes to the one loop determinant 
 when $n \in D$. 
For these values of $n$, the 
admissible solution, near the boundary of $AdS_{2}$, for $f(z)$ is $S_{2+}(z)$, whereas for $(b(z),c(z))$ system, the valid solution is $S_{2-}(z)$. Furthermore, we will see that for $n\in D$, 
 the variation of the logarithm of one loop determinant is no
  longer total derivative.  It contains a bulk term together with a boundary term. \\
 Let us consider the case with $p>0$. In this case we construct the 
 bosonic Green's function with the mode $S_{1+}(z)$ and $S_{2+}(z)$ and the 
 fermionic Green's function with the mode $S_{1+}(z)$ and $S_{2-}(z)$. 
 Repeating the previous analysis, we find that the 
 contribution to the variation of one loop determinant 
 in the range $\frac{\Delta-1}{2L}<n<\frac{\Delta}{2L}$ and for $p>0$ is
\bea\label{eq:bulk.AdS2.1}
\frac{\delta}{\delta\alpha}\ln \hat Z^{+}_{1-loop}(\a)&=&-\frac{i L q S_{2-}(z) ((2 p (-1 + z) + (2 L n - \Delta) z) S_{1+}(z) + 4 (-1 + z) z \p_{z}S_{1+}(z))}{2 \sqrt{1 - z} (2 p - \Delta + 2 L (n + iq \alpha)) c_{1+-}}\Big|^{1}_{0}\,\nn\\
&&+\text{Bulk term}\,,
\eea
where the bulk term is given  by
\bea\label{eq:bulk.AdS2.1.1}
\text{Bulk term}&=&-\int^{1}_{0}dz\,\frac{ L q (-i + 2 L q \a) S_{1+}(z) S_{2-}(z)}{4c_{1+-} \sqrt{1- z}}\nn\\
&&+\int^{1}_{0}dz\,\frac{ L q (-i + 2 L q \a) S_{1+}(z) S_{2+}(z)}{4c_{1++} \sqrt{1- z}}\,.
\eea
In the above expression for the bulk term, the first term comes from the  trace of the
fermionic Green's function while the second term is the trace of the bosonic Green's function.
Similarly the contribution to the variation of one loop determinant in the range $\frac{\Delta-1}{2L}<n<\frac{\Delta}{2L}$ and for $p\leq0$ is
\bea\label{eq:bulk.AdS2.2}
\frac{\delta}{\delta\alpha}\ln \hat Z^{-}_{1-loop}(\a)&=&-\frac{i L q S_{2-}(z) ((2 p (-1 + z) + (2 L n - \Delta) z) S_{1-}(z) + 4 (-1 + z) z \p_{z}S_{1-}(z))}{2 \sqrt{1 - z} (2 p - \Delta + 2 L (n + iq \alpha)) c_{1--}}\Big|^{1}_{0}\,\nn\\
&&+\text{Bulk term}\,,
\eea
where now the bulk term is given as
\bea\label{eq:bulk.AdS2.2.1}
\text{Bulk term}&=&-\int^{1}_{0}dz\,\frac{ L q (-i + 2 L q \a) S_{1-}(z) S_{2-}(z)}{4c_{1--} \sqrt{1- z}}\nn\\
&&+\int^{1}_{0}dz\,\frac{ L q (-i + 2 L q \a) S_{1-}(z) S_{2+}(z)}{4c_{1-+} \sqrt{1- z}}\,.
\eea

The contributions from the boundary terms in \eqref{eq:bulk.AdS2.1} and \eqref{eq:bulk.AdS2.2} are given in \eqref{eq:bdyterm.3} and \eqref{eq:bdyterm.4}, respectively. 
Note that the contributions from the boundary terms are easy to integrate with respect to 
$\alpha$ and result in logarithims to the one loop free energy. 

Let us proceed to evaluate the bulk terms \eqref{eq:bulk.AdS2.1.1} and \eqref{eq:bulk.AdS2.2.1}. 
These terms 
 involve an integration of a product of Hypergeometric functions.  The relevant 
 integrals are listed in appendix \ref{appendixc}.  The results are linear combinations of 
 digamma functions.
After this we need to further integrate these terms with respect to $\alpha$. This is reasonably simple 
to do, since the digamma functions are derivatives of the logarithm of the gamma functions 
and the variable $\alpha$ occurs linearly in  their arguments.  

\noindent
{\bf One loop determinant:}
Assimilating all the contributions of all the Kaluza-Klein modes $n$, keeping track 
of the sign of $p$ and performing the integration with respect to $\alpha$ we obtain 
the following answer of the one loop determinant.
\begin{eqnarray} \label{mainresult.1}
\ln Z(\a,\Delta)&=&\sum_{p>0,n>\frac{\Delta}{2L}}\ln\Big(p+L(n+iq\a)-\frac{\Delta}{2}\Big)-\sum_{p\leq0,n<\frac{\Delta-1}{2L}}\ln\Big(-p-L(n+iq\a)+\frac{\Delta}{2}\Big)\nn\\
&&-\sum_{\stackrel{p\leq0}{\frac{\Delta-1}{2L}<n<\frac{\Delta}{2L}}}\ln\Big(-p-L(n+iq\a)+\frac{\Delta}{2}\Big)-\sum_{\frac{\Delta-1}{2L}<n<\frac{\Delta}{2L}}\sum_{p\in\mathbb Z}\ln\frac{\Gamma(\frac{1}{2}+\frac{1}{4}\hat x)\Gamma(\frac{1}{4}\hat x^{*})}{\Gamma(\frac{1}{2}+\frac{1}{4}\hat y)\Gamma(1+\frac{1}{4}\hat y^{*})}\,,\nn\\
\end{eqnarray}
where 
\be
\hat x=2|p|+\Delta-2Ln+2iLq\a\,,\quad \hat y=2|p|-\Delta+2Ln-2iLq\a\,.
\ee
The first line  in (\ref{mainresult.1}) is the total contribution from the boundary terms for the 
Kaluza-Klein modes in the 
range $n>\Delta/2L$ and $n < ( \Delta-1)/2L$. 
The second line contains the contribution of the bulk term as well as the boundary term in 
  (\ref{eq:bulk.AdS2.1})  and (\ref{eq:bulk.AdS2.2})
  when $n \in D$.  The bulk integral results in  product of Gamma functions.
  We note that (\ref{mainresult.1}) is not yet final result as we still need to include contributions from KK modes for which $\frac{\Delta}{2L}$ or $\frac{\Delta-1}{2L}$ or both are integer. However, if $\Delta$ and $L$ are such that none of these ratios are integers then (\ref{mainresult.1}) is the final result. As we noted earlier when such ratios are integer the modes are at the border of normalizabilty and therefore, it is not very clear how to take into account their contributions to partition function which is computed with normalizable boundary conditions. In this scenario we take a clue from the free theory partition function obtained in appendix \ref{appendixb}, that is that the partition function is a continuous function of $\Delta$. We assume that this holds true even in the presence of non zero charge $q$ {\it{i.e.}} we require that $\ln Z(\a,\Delta)$ is continuous across every real value of $\Delta$. With this requirement we find that the partition function is\footnote{From (\ref{mainresult.1}) we find that for any $\Delta_{0}$, in particular for $\Delta_{0}=1\,\,\text{and}\,\, 0$, $\lim_{\epsilon\rightarrow 0}\ln Z(\a,\Delta_{0}-\epsilon)=\lim_{\epsilon\rightarrow 0}\ln Z(\a,\Delta_{0}+\epsilon)$. We arrive at (\ref{mainresult}) by requiring that $\lim_{\epsilon\rightarrow 0}\ln Z(\a,\Delta_{0}-\epsilon)=\lim_{\epsilon\rightarrow 0}\ln Z(\a,\Delta_{0}+\epsilon)=\ln Z(\a,\Delta_{0})$.}
  \begin{eqnarray} \label{mainresult}
\boxed{
\begin{aligned}
\ln Z(\a,\Delta)&=\sum_{p>0,n\geq\ceil{\frac{\Delta}{2L}}}\ln\Big(p+L(n+iq\a)-\frac{\Delta}{2}\Big)-\sum_{p\leq0,n<\frac{\Delta-1}{2L}}\ln\Big(-p-L(n+iq\a)+\frac{\Delta}{2}\Big)\nn\\
&-\sum_{\stackrel{p\leq0}{\ceil{\frac{\Delta-1}{2L}}\leq n<\frac{\Delta}{2L}}}\ln\Big(-p-L(n+iq\a)+\frac{\Delta}{2}\Big)\nn\\
&-\sum_{{\frac{\Delta-1}{2L}}< n<\frac{\Delta}{2L}}\sum_{p\in\mathbb Z}\ln\frac{\Gamma(\frac{1}{2}+\frac{1}{4}\hat x)\Gamma(\frac{1}{4}\hat x^{*})}{\Gamma(\frac{1}{2}+\frac{1}{4}\hat y)\Gamma(1+\frac{1}{4}\hat y^{*})}\,.
\end{aligned}
}\\
\end{eqnarray}
The ceiling function $\ceil{x}$ in the above sum gives an integer greater than or equal to $x$.
  The equation (\ref{mainresult}) is  our final result for the one loop determinant 
  for a chiral multiplet coupled to a  supersymmetric background vector multiplet
  on $AdS_2\times S^1$. Let us recall that we have used normalizable boundary conditions
in  $AdS_2$   and periodic boundary conditions for the fermions on $S^1$. \\
There  could be  an $\alpha$ independent constant since we have 
obtained the result in (\ref{mainresult}) by first differentiating the one loop determinant
with respect $\alpha$, evaluate this using Green's function approach and then integrating 
back with respect to $\alpha$.  This constant could depend on both $L $ and $\Delta$. 
 To study the $\Delta$ dependence, we can repeat the above steps but now taking variation with respect to $\Delta$. It turns out that  in the 
range $n>\frac{\Delta}{2L}$ and $n < \frac{ \Delta-1}{2L}$ where there are no bulk terms, all the steps go through as before, bulk terms cancel  and one arrives at boundary terms. However, in this case the boundary term receives extra contributions in addition to the first line in (\ref{mainresult.1}). These are terms which are independent of $p$ and therefore, vanish after 
summing $p$ from $-\infty$ to $\infty$ by using zeta function regularization.
 Note that the variations with respect to $\alpha$ and $\Delta$ are integrable and fix the one loop determinant upto an overall $L$ dependent constant. 
In the range  $\frac{ \Delta-1}{2L} < n <\frac{\Delta}{2L}$, the integral appearing in $\text{Tr}[G_{F}\frac{\delta}{\delta \Delta}]$ is divergent near $z=1$ (i.e. near the boundary of $AdS_2$). 
It is possible to regularize properly and compute these variations. We find that bulk terms agree as well with the result obtained from the $\a$ variation. 
We will now show by comparison with explicit eigen function calculations 
for $\alpha =0$,  $L=1, L=2$ and $0< \Delta <2$ that  (\ref{mainresult}) precisely 
agrees with the one loop determinant obtained by the eigen function method.

\noindent
{\bf  Comparison with eigen function result:    $L=1$,  $0< \Delta <2$.}

When $q =0$, the standard action in (\ref{eq:susyS.1}) reduces to that of the
free boson and free fermion in $AdS_2\times S^1$.   The background vector multiplet 
decouples. 
The eigen function method 
has been applied to evaluate partition functions for such actions by  \cite{Klebanov:2011uf}. 
We use this approach and evaluate the partition function 
for the chiral multiplet with $q=0, L=1$ and  $\Delta$ in the range
$0<\Delta <2$. This has been done in appendix \ref{appendixb}. The result is given 
in (\ref{lequal1a}) for $0<\Delta <1$ and (\ref{lequal1b}) for $1<\Delta<2$. 
Note that these expressions  are symmetric under the transformation 
$\Delta \rightarrow 2-\Delta$. This symmetry is manifest in  the  
integral representation for the free energy given in (\ref{oneloopint})
\footnote{In fact from the integral representation of the free energy (\ref{oneloopint})
we see that there are the following symmetries at $q=0$: $\Delta \rightarrow \Delta + 2L$ for arbitrary $L$.
For $L = \frac{1}{N}$, where $N$ is a positive integer, we have  the symmetry
$\Delta \rightarrow \frac{2}{N} -\Delta$.}.
It can also be verified easily  by examining the partition function in the infinite
product form given in (\ref{lequal1a}) and (\ref{lequal1b}). 

We will  verify  the  general result for the  partition function given in (\ref{mainresult}) 
agrees with that obtained by the eigen function method for $L=1$ for $q=0$. 
First let us write down the expression in (\ref{mainresult}) for $L=1$ and 
$1<\Delta<2$.  Note that  for this situation, there is no contribution 
from the second line of (\ref{mainresult}). 
Thus we obtain 
\begin{eqnarray} \label{greater}
\ln Z(\a)|_{L=1, 1< \Delta<2} 
= \sum_{p =1, n = 1}^\infty \ln( p + n + i q\alpha - \frac{\Delta}{2} ) 
- \sum_{p =0, n =0}^\infty \ln ( p + n - i q\alpha + \frac{\Delta}{2} ) \,.
\end{eqnarray}
Now for $L=1$ and $0<\Delta<1$, the  only integer  allowed in the domain $D$ is 
$n=0$, the second line  in (\ref{mainresult}) contributes. 
We can  therefore take
\begin{equation}
\hat x|_{n=0, L=1} = 2 |p|  +\Delta  + 2i q \alpha, 
\qquad \hat y|_{ n=0, L=1} = 2|p|  -\Delta   - 2i q\alpha\,.
\end{equation}
We expand the term involving the gamma functions using the 
identity
\begin{equation}
\Gamma(z) = \frac{e^{\gamma z}}{z} \prod_{n=1}^\infty \left( 1+ \frac{z}{n} \right)^{-1} 
e^{\frac{z}{n}}\,.
\end{equation}
This leads to 
\begin{eqnarray}\label{congamma}
{\cal I} &=&\left.  \sum_{p \in \mathbb{Z}} \ln \frac{ \Gamma(  \frac{1}{2} + \frac{\hat x}{4} ) 
 \Gamma(  \frac{\hat x^*}{4} ) }{
 \Gamma(  \frac{1}{2} + \frac{\hat y}{4} ) 
 \Gamma( 1+ \frac{\hat y^*}{4} ) }\right|_{n = 0, L=1}\,, \\ \nonumber
 &=& - \sum_{n=0, p=1}^\infty 2 \ln ( n + p + \frac{\Delta}{2} + i  q\alpha) 
 - \sum_{n=0}^\infty \ln ( n + \frac{\Delta}{2} + i q \alpha)  \\ \nonumber
& &  +  \sum_{n=1, p=1}^\infty  2\ln ( n + p -\frac{\Delta}{2} - i q\alpha) 
 + \sum_{n=1}^\infty \ln( n - \frac{\Delta}{2} - i q\alpha )  \\ \nonumber
 & & - \sum_{p \in \mathbb{Z} } 
 \left[ \ln \left( \frac{ |p| + \frac{\Delta}{2} - i q \alpha}{ |p| + \frac{\Delta}{2} + i q \alpha} \right) 
 - \sum_{n=1}^\infty \ln \left(  \frac{ 2n + |p| + \frac{\Delta}{2} - i q \alpha}{
 2n + |p| + \frac{\Delta}{2} + i q\alpha} \right) \right] \\ \nonumber
 & & - \sum_{p \in \mathbb{Z} } 
 \left[ \ln \left( \frac{ 2+  |p| + \frac{\Delta}{2} - i q \alpha}{ 2+  |p| + \frac{\Delta}{2} + i q \alpha} \right) 
 - \sum_{n=1}^\infty \ln \left(  \frac{ 2(n+1)  + |p| + \frac{\Delta}{2} - i q \alpha}{
 2(n+1)  + |p| + \frac{\Delta}{2} + i q\alpha} \right) \right]\,. \\ \nonumber
\end{eqnarray}
To obtain this we have used $\zeta (0) = -\frac{1}{2} $.
Combining this with the rest of the terms of (\ref{mainresult}) we obtain 
\begin{eqnarray}\label{partman}
\ln Z(\a)|_{L=1, 0<\Delta <1} &=& 
\sum_{n=0, p=1}^\infty \ln ( p + n + \frac{\Delta}{2} + i q \alpha) \nn
 \\ \nonumber
&& +  \sum_{n=1, p=1}^\infty 
\ln \left( \frac{ p + n + \frac{\Delta}{2} + i q\alpha}{p + n + \frac{\Delta}{2} - i q\alpha} \right) 
+ \sum_{n=1, p=-\infty}^\infty \ln \left( 
\frac{ 2n + |p| + \frac{\Delta}{2} - i q\alpha} { 2n + |p| + \frac{\Delta}{2} + i q\alpha} 
  \right) \\ \nonumber
  & & - \sum_{n=0, p=1}^\infty \ln ( n +  p- \frac{\Delta}{2} - i q\alpha)    
  \\  \nonumber
  &&  + \sum_{n=1, p =1}^\infty 
  \ln \left( \frac{ p + n - \frac{\Delta}{2} + i q\alpha}{p + n- \frac{\Delta}{2} - i q\alpha} \right) 
  +
  \sum_{n=1, p = -\infty}^\infty 
  \ln \left( 
\frac{ 2n + |p| - \frac{\Delta}{2} - i q\alpha} { 2n + |p| + \frac{\Delta}{2} + i q\alpha} 
  \right)\,.\\
\end{eqnarray}
Note that on re-organising the two terms on the second line  of (\ref{partman}) 
we see that they combine and cancel each other. The same property  holds also for the 
two terms of the fourth line of (\ref{partman}). 
Therefore we are left with 
\begin{eqnarray}\label{lesser}
\ln Z(\a)|_{L=1, 0<\Delta <1} &=& 
\sum_{n=0, p=1}^\infty \ln ( p + n + \frac{\Delta}{2} + i q \alpha)  
 - \sum_{n=0, p=1}^\infty \ln ( n +  p- \frac{\Delta}{2} - i q\alpha)\,.\nn\\ 
\end{eqnarray}
We can now see that replacing $\Delta \rightarrow 2-\Delta$ in  (\ref{greater}) we obtain 
(\ref{lesser}).  Thus this symmetry which observed in the partition function 
at $L=1, q=0$  in (\ref{oneloopint}) continues to hold when $q\neq 0$. 
Furthermore note that on substituting $q=0$ in (\ref{greater}) and (\ref{lesser}) 
it precisely agrees with the  eigen function partition function in (\ref{lequal1a}) and (\ref{lequal1b}). 
This agreement and the fact that (\ref{mainresult}) has the $\Delta \rightarrow 2-\Delta$ 
symmetry for $L=1$ serves as a non-trivial check of the Green's function method. 
This also implies that the  $L$ dependent integration constant  in 
(\ref{mainresult}) is zero at least for $L=1$. \\
We have also verified that the partition function (\ref{mainresult}) has the symmetry $\Delta \rightarrow 2L-\Delta$ for $L=\frac{1}{2}$ and non zero $q$. Given these observations for $L=1$ and $L=\frac{1}{2}$, it seems likely that the partition function is invariant under $\Delta \rightarrow 2L-\Delta$ for $L=\frac{1}{N}$ and for non zero $q$, although it will be very nice to prove it in general. Moreover, it is easy to see that the partition function (\ref{mainresult}) is also invariant under $\Delta \rightarrow 2L+\Delta$ for non zero $q$ and arbitrary $L$. Thus this symmetry also continues to hold for non zero $q$.

\noindent
{\bf  Comparison with the eigen function result:    $L=2$,  $0< \Delta <2$.}

As  a further check on the Green's function method we compare (\ref{mainresult}) 
at $q=0$ with the eigen function result for $L=2$. 
First consider the domain $1<\Delta<2$.  In this interval there is no
contribution from the second line of (\ref{mainresult}). 
The first line yields
\begin{eqnarray}
\ln Z|_{L = 2, 1<\Delta<2} =  \sum_{p=1, n=1}^\infty \ln ( p + 2n - \frac{\Delta}{2} ) 
- \sum_{p=0, n=0}^\infty \ln ( p + 2n + \frac{\Delta}{2} ) \,.
\end{eqnarray} 
This expression precisely agrees with that obtained by the eigen functionmethod 
in (\ref{l2a}). 
Now lets examine the domain $0<\Delta<1$. The integer allowed in the domain 
$D$ is $n=0$, thus now the second line of (\ref{mainresult}) contributes. 
Note however since only $n=0$  contributes, the values of $\hat x$ and $\hat y$ is 
independent of $L$ and therefore  the contribution of the term involving 
the gamma function is same as that given in (\ref{congamma}) with $q =0$. 
Taking all this into account we find  for $0<\Delta<1$, and $L=2$, 
(\ref{mainresult}) reduces to 
\begin{eqnarray}
\ln Z|_{L = 2, 0<\Delta< 1} &=&  \sum_{p=1, n=1}^\infty \ln ( p + 2n - \frac{\Delta}{2} )  
- \sum_{p=0, n=1}^\infty \ln ( p + 2n + \frac{\Delta}{2} ) 
\\ \nonumber
& & +\sum_{n=0, p=1}^\infty 2 \ln ( n + p + \frac{\Delta}{2} )  \\ \nonumber
& & -\sum_{n=1, p=1}^\infty  2\ln ( n + p - \frac{\Delta}{2})  
- \sum_{n=1}^\infty\ln ( n - \frac{\Delta}{2})\,.
\end{eqnarray}
Comparing the above equation and the result from the eigen function method in  (\ref{l2b}) and 
after some obvious rearrangement of terms we see that they precisely agree. 
The agreement of the result (\ref{mainresult}) with the eigen function method for $L=2, q=0$ in domain
$0<\Delta<2$ serves as another check of the green function method developed in this paper. 
Note again, the fact that we have obtained agreement with the eigen function method 
for $L=2$ implies that the the putative $L$ dependent integration constant 
in (\ref{mainresult})  is zero.

\subsection{One loop determinant of the $Q$-exact action}

In this section we repeat the above analysis for the Q-exact deformations presented in the section \ref{Sec:Q-exact.1}. In order to compute the Green's function, we need to find the solutions of the differential equations \eqref{eq:f.AdS2.2} and \eqref{eq:c.AdS2.2}. 
These differential equations are certainly different from that of the equations of motion 
for $f$ and $(b, c)$ obtained from the standard action which are given in 
(\ref{eq:f.AdS2.1}),(\ref{eq:c.AdS2.1}).  
However it can be easily seen that the 
asymptotic properties  of the solutions as $z\rightarrow 1$  to these fields 
for both the $Q$ exact action as well the standard action are the same. 
Also the equation relating the field $b$ to $c$ in (\ref{eq:b.AdS2.1}) and (\ref{eq:b.AdS2.2})\,.
Therefore the asymptotic behaviour of the fields are given by 
(\ref{bcfnp}), (\ref{assce}) and (\ref{bccb}). 
As mentioned earlier we need to impose 
normalizable boundary conditions on all the fields to ensure that the path integral 
is well defined. Thus the normalizable solution for all the Kaluza-Klein modes are 
given by  (\ref{finalnorm})\,.

Let us now proceed to solve the equations \eqref{eq:f.AdS2.2} and \eqref{eq:c.AdS2.2} and
 construct the Green's function. 
 We will implement smoothness near $z=0$ and normalizable boundary conditions at 
 $z=1$. 
 The solution for  this equation 
 which is smooth near $z=0$ for $p>0$ is given by 
\be
\wt S_{1+}(z)=(1 - z)^{\frac{1}{4}( - 2 L n + \Delta)} z^{p/2}\,{}_2F_{1}[a_{1}, b_{1}, 1 + p, z]\,,
\ee
whereas the solution which is smooth near $z=0$ for $p<0$
\be
\wt S_{1-}(z)=(1 - z)^{\frac{1}{4}( - 2 L n + \Delta)} z^{-p/2} \, {}_2F_{1}[ \wt a_{1},\wt b_{1}, 1 - p, z] \,.
\ee
Here
\bea
&&a_{1}= \frac{1}{4}(1 - 2 L n + 2 p + \Delta -\sqrt{1-4Ln-4p+2\Delta-4L^{2}q^{2}\a^{2}})\,,\nn\\
&&b_{1}=\frac{1}{4}(1 - 2 L n + 2 p + \Delta +\sqrt{1-4Ln-4p+2\Delta-4L^{2}q^{2}\a^{2}})\,,\nn\\
&&\wt a_{1}= \frac{1}{4}(1 - 2 L n - 2 p + \Delta -\sqrt{1-4Ln-4p+2\Delta-4L^{2}q^{2}\a^{2}})\,,\nn\\
&&\wt b_{1}=\frac{1}{4}(1 - 2 L n - 2 p + \Delta +\sqrt{1-4Ln-4p+2\Delta-4L^{2}q^{2}\a^{2}})\,.
\eea
For $p=0$, both the solution coincide. The second solution for $p=0$ is logarithmic in $z$ and therefore will not consider in the analysis. Next we look for the solution near $z=1$. The 
normalizable 
solution for $n>\frac{\Delta-1}{2L}$ is
\be
\wt S_{2+}(z)=(1 - z)^{\frac{1}{4}(2 + 2 L n - \Delta)} z^{p/2}\,{}_2F_{1}[a_{2}, b_{2}, \frac{3}{2}+Ln-\frac{\Delta}{2}, 1-z]\,,
\ee
whereas the admissible solution for $n<\frac{\Delta-1}{2L}$ is
\be
\wt S_{2-}(z)=(1 - z)^{\frac{1}{4}(-2 L n+\Delta)} z^{p/2}\,{}_2F_{1}[\wt a_{2},\wt b_{2}, \frac{1}{2}-Ln+\frac{\Delta}{2}, 1-z]\,.
\ee
Here
\bea\label{eq:AdS2.HyperCoeff}
&&a_{2}=\frac{1}{4}(3+2Ln+2p-\Delta-\sqrt{1-4Ln-4p+2\Delta-4L^{2}q^{2}\a^{2}})\,, \nn\\
&&b_{2}=\frac{1}{4}(3+2Ln+2p-\Delta+\sqrt{1-4Ln-4p+2\Delta-4L^{2}q^{2}\a^{2}})\,,\nn\\
&&\wt a_{2}=\frac{1}{4}(1-2Ln+2p+\Delta-\sqrt{1-4Ln-4p+2\Delta-4L^{2}q^{2}\a^{2}})\,,\nn\\
&&\wt b_{2}=\frac{1}{4}(1-2Ln+2p+\Delta+\sqrt{1-4Ln-4p+2\Delta-4L^{2}q^{2}\a^{2}})\,.
\eea
For $n=\frac{\Delta-1}{2L}$, both the solutions coincide and goes like $(1-z)^{1/4}$ which is at the border of normalisability. The other solution is logarithmic in $(1-z)$ near $z=1$ which is not square integrable.\\
Analysing the $(b,c)$ system we find the similar situation as in the previous case. For the $(b,c)$ system, we have the admissible solution $\wt S_{2+}(z)$ for $n>\frac{\Delta}{2L}$ and $\wt S_{2-}(z)$ is admissible solution for $n<\frac{\Delta}{2L}$. So we see that for $n>\frac{\Delta}{2L}$ and $n<\frac{\Delta-1}{2L}$ both $f(z)$ and $(b(z),c(z))$ have the same admissible solutions. However there is mismatch of the solution in the range $\frac{\Delta-1}{2L}<n<\frac{\Delta}{2L}$. \\
At $n=\frac{\Delta}{2L}$ (if it is an integer) we have the similar feature as explained in the previous section. If we begin with $c_{\frac{\Delta}{2L},p}(z)=\wt S_{2+}(z)$ which goes like $\sqrt{1-z}$ as $z\rightarrow 1$, the corresponding mode for $b_{\frac{\Delta}{2L},p}(z)$ goes like $\mathcal O(1)$ and is at the border of normalizability. On the other hand for $c_{\frac{\Delta}{2L},p}(z)=\wt S_{2-}(z)$ which goes like $\mathcal O(1)$ as $z\rightarrow 1$, the corresponding mode for $b_{\frac{\Delta}{2L},p}(z)$ goes like $\sqrt{1-z}$ and is admissible. Thus for the $(b,c)$ system none of the modes at $n=\frac{\Delta}{2L}$ satisfy strict normalizable boundary conditions.

\noindent
{\bf{Case 2:  $n> \frac{\Delta}{2L}$ and $n<\frac{\Delta-1}{2L}$ }}\\
Proceeding as before we find that for $n>\frac{\Delta}{2L}$ and $n<\frac{\Delta-1}{2L}$, the variation of the one loop determinant is again total derivative and is given by
\be\label{eq:Bdy.AdS2.2}
\frac{\delta}{\delta\alpha}\ln \wt Z_{1-loop}(\a)=-\frac{i L q \wt S_{2}(z) ((2 p (-1 + z) + (2 L n - \Delta) z) \wt S_{1}(z) + 4 (-1 + z) z \p_{z}\wt S_{1}(z))}{2 \sqrt{1 - z} (2 p - \Delta + 2 L (n + iq \alpha)) \wt c_{1}}\Big|^{1}_{0}\,.
\ee
Here $\wt c_{1}$ is a constant determined by evaluating the Wronskian near $z=0$. For various range of $(n,p)$, this constant is given as follows
\bea
&&\wt c_{1++} = -p \frac{\Gamma(p) \Gamma(\frac{3}{2} + L n - \frac{\Delta}{2})}{\Gamma(a_{2}) \Gamma(b_{2})},\,\,\quad\quad \text{for }\,\,p>0, n>\frac{\Delta}{2L}\,,\nn\\
&&\wt c_{1-+}=p \frac{\Gamma(-p) \Gamma(\frac{3}{2} + L n - \frac{\Delta}{2})}{\Gamma(a_{2}-p) \Gamma(b_{2}-p)},\,\,\quad\quad\text{for }\,\,p\leq0, n>\frac{\Delta}{2L}\,,\nn\\
&&\wt c_{1+-} = -p \frac{\Gamma(p) \Gamma(\frac{1}{2}(1 -2 L n +\Delta)}{\Gamma(\wt a_{2}) \Gamma(\wt b_{2})},\,\quad \text{for }\,\,p>0, n<\frac{\Delta-1}{2L}\,,\nn\\
&&\wt c_{1--}=p\frac{\Gamma(-p) \Gamma(\frac{1}{2}(1 -2 L n +\Delta))}{\Gamma(\wt a_{1}-p) \Gamma(\wt b_{2}-p)},\,\quad\text{for }\,\,p\leq0, n<\frac{\Delta-1}{2L}.
\eea
Note that though there seems to be an explicit dependence on $\wt c_1$  in denominator of 
 the boundary terms
(\ref{eq:Bdy.AdS2.2}), the numerator also  depends linearly  on $\wt c_1$  when we substitute
the behaviour of the functions at $0$ and $1$. In fact for the  Kaluza-Klein modes
$n>\frac{\Delta}{2L}$ and $n<\frac{\Delta-1}{2L}$ we do not need the precise 
knowledge of the normalisation constant $\wt c_1$. The final result for the boundary terms is independent of the 
constant $\wt c_1$. 
We now evaluate the right hand side of \eqref{eq:Bdy.AdS2.2}, which we denote below by $\wt{\text{BT}}$, for various range of $(n,p)$.\\
${n>\frac{\Delta}{2L}}$ and ${p>0}:$ In this case we have $ S_{1}(z)=\wt S_{1+}(z)$ and $S_{2}(z)=\wt S_{2+}(z)$. Therefore, the boundary term is 
\be\label{eq:wtbdyterm.1}
\wt{\text{BT}}=\frac{2iLq}{2p-\Delta+2L(n+iq\alpha)}\,.
\ee
${n>\frac{\Delta}{2L}}$ and ${p\leq0}:$ In this case we have $S_{1}(z)=\wt S_{1-}(z)$ and $S_{2}(z)=\wt S_{2+}(z)$. Therefore, the boundary term is 
\be\label{eq:wtbdyterm.2}
\wt{\text{BT}}=0\,.
\ee
${n<\frac{\Delta-1}{2L}}$ and ${p>0}:$ In this case we have $S_{1}(z)=\wt S_{1+}(z)$ and $S_{2}(z)=\wt S_{2-}(z)$. Therefore, the boundary term is 
\be\label{eq:wtbdyterm.3}
\wt{\text{BT}}=0\,.
\ee
${n<\frac{\Delta-1}{2L}}$ and ${p\leq0}:$ In this case we have $S_{1}(z)=\wt S_{1-}(z)$ and $S_{2}(z)=\wt S_{2-}(z)$. Therefore, the boundary term is 
\be\label{eq:wtbdyterm.4}
\wt{\text{BT}}=-\frac{2iLq}{2p-\Delta+2L(n+iq\alpha)}\,.
\ee

\noindent
{\bf{Case 2:  $\frac{\Delta-1}{2L}<n<\frac{\Delta}{2L}$}}\\
As in the previous case of 
the standard action, 
 we also need to consider the contribution to the one loop determinant for $\frac{\Delta-1}{2L}<n<\frac{\Delta}{2L}$. We begin with the case for $p>0$. In this case we construct the bosonic Green's function with the mode $\wt S_{1+}(z)$ and $\wt S_{2+}(z)$ and the fermionic Green's function with the mode $\wt S_{1+}(z)$ and $\wt S_{2-}(z)$. Repeating the previous analysis, we find that the contribution to the variation of one loop determinant in this range is
\bea\label{eq:bulk.AdS2.3}
\frac{\delta}{\delta\alpha}\ln \hat Z^{+}_{1-loop}(\a)&=&-\frac{i L q \wt S_{2-}(z) ((2 p (-1 + z) + (2 L n - \Delta) z) \wt S_{1+}(z) + 4 (-1 + z) z \p_{z}\wt S_{1+}(z))}{2 \sqrt{1 - z} (2 p - \Delta + 2 L (n + iq \alpha)) \wt c_{1+-}}\Big|^{1}_{0}\,\nn\\&&+\text{Bulk term}
\eea
where the bulk term is given as
\bea\label{eq:bulk.AdS2.1.2}
\text{Bulk term}&=&-\int^{1}_{0}dz\,\frac{ L^{2} q^{2}\a  \wt S_{1+}(z) \wt S_{2-}(z)}{2\wt c_{1+-} \sqrt{1- z}}+\int^{1}_{0}dz\,\frac{ L^{2} q^{2} \a\wt S_{1+}(z) \wt S_{2+}(z)}{2\wt c_{1++} \sqrt{1- z}}\,.\nn\\
\eea
In the above expression, the first term comes from the trace of the  fermionic Green's function and the second term is the trace of the bosonic Green's function.
Similarly the contribution to the variation of one loop determinant in the range $\frac{\Delta-1}{2L}<n<\frac{\Delta}{2L}$ and for $p\leq0$ is
\bea\label{eq:bulk.AdS2.4}
\frac{\delta}{\delta\alpha}\ln \hat Z^{-}_{1-loop}&=&-\frac{i L q \wt S_{2-}(z) ((2 p (-1 + z) + (2 L n - \Delta) z) \wt S_{1-}(z) + 4 (-1 + z) z \p_{z}\wt S_{1-}(z))}{2 \sqrt{1 - z} (2 p - \Delta + 2 L (n + iq \alpha)) \wt c_{1--}}\Big|^{1}_{0}\,\nn\\
&&+\text{Bulk term}\,,
\eea
where now the bulk term is given as
\bea\label{eq:bulk.AdS2.2.2}
\text{Bulk term}&=&-\int^{1}_{0}dz\,\frac{ L ^{2}q^{2} \a \wt S_{1-}(z) \wt S_{2-}(z)}{2\wt c_{1--} \sqrt{1- z}}+\int^{1}_{0}dz\,\frac{ L^{2} q^{2} \a \wt S_{1-}(z) \wt S_{2+}(z)}{2\wt c_{1-+} \sqrt{1- z}}\,.\nn\\
\eea
The contributions from the boundary terms in \eqref{eq:bulk.AdS2.3} and \eqref{eq:bulk.AdS2.4} are given in \eqref{eq:wtbdyterm.3} and \eqref{eq:wtbdyterm.4}, respectively. 
Next we would like to evaluate the bulk terms \eqref{eq:bulk.AdS2.1.2} and \eqref{eq:bulk.AdS2.2.2} which involve an integration of a product of Hypergeometric functions.
The result of these integrations again involve linear combination of digamma functions. 
They are listed in the appendix \ref{appendixc}.
Then further integrating with respect to $\alpha$ and assimilating all the contributions
we obtain the following result for the one loop determinant for the $Q$-exact localising
action. 
\bea\label{mainresult2}
\ln Z(\a,\Delta)&=&\sum_{p>0,n>\frac{\Delta}{2L}}\ln\Big(p+L(n+iq\a)-\frac{\Delta}{2}\Big)-\sum_{p\leq0,n<\frac{\Delta-1}{2L}}\ln\Big(-p-L(n+iq\a)+\frac{\Delta}{2}\Big)\nn\\
&&-\sum_{p\leq0,\frac{\Delta-1}{2L}<n<\frac{\Delta}{2L}}\ln\Big(-p-L(n+iq\a)+\frac{\Delta}{2}\Big)-\sum_{\frac{\Delta-1}{2L}<n<\frac{\Delta}{2L}}\sum_{p>0}\ln\frac{\Gamma(\wt a_{2})\Gamma(b_{2})}{\Gamma(\wt b_{2})\Gamma(a_{2})}\nn\\
&&-\sum_{\frac{\Delta-1}{2L}<n<\frac{\Delta}{2L}}\sum_{p\leq0}\ln\frac{\Gamma(\wt a_{2}-p)\Gamma(b_{2}-p)}{\Gamma(\wt b_{2}-p)\Gamma(a_{2}-p)}\,.
\eea
As discussed in the previous section, the above expression for the partition function is not yet the final result for the $Q$ exact deformation. We still need to include contributions from KK modes for which $\frac{\Delta}{2L}$ or $\frac{\Delta-1}{2L}$ or both are integer. However, if $\Delta$ and $L$ are such that these ratios are not integers then (\ref{mainresult2}) is the final result for the $Q$-exact deformation. As we noted earlier when such ratios are integer the modes are at the border of normalizabilty and therefore, it is not very clear how to take into account their contributions to partition function which is computed with normalizable boundary conditions. Following the previous analysis we also assume here that the partition function is a continuous as a function of $\Delta$ even in the presence of non zero charge $q$ {\it{i.e.}} we require that $\ln Z(\a,\Delta)$ is continuous across every real value of $\Delta$. With this requirement we find that the partition function is
\bea\label{mainresult2.2}
\ln Z(\a,\Delta)&=&\sum_{p>0,n\geq\ceil{\frac{\Delta}{2L}}}\ln\Big(p+L(n+iq\a)-\frac{\Delta}{2}\Big)-\sum_{p\leq0,n<\frac{\Delta-1}{2L}}\ln\Big(-p-L(n+iq\a)+\frac{\Delta}{2}\Big)\nn\\
&&-\sum_{p\leq0,\ceil{\frac{\Delta-1}{2L}}\leq n<\frac{\Delta}{2L}}\ln\Big(-p-L(n+iq\a)+\frac{\Delta}{2}\Big)-\sum_{{\frac{\Delta-1}{2L}}< n<\frac{\Delta}{2L}}\sum_{p>0}\ln\frac{\Gamma(\wt a_{2})\Gamma(b_{2})}{\Gamma(\wt b_{2})\Gamma(a_{2})}\nn\\
&&-\sum_{{\frac{\Delta-1}{2L}}< n<\frac{\Delta}{2L}}\sum_{p\leq0}\ln\frac{\Gamma(\wt a_{2}-p)\Gamma(b_{2}-p)}{\Gamma(\wt b_{2}-p)\Gamma(a_{2}-p)}\,.
\eea

Let us compare the result for the one loop determinant of the $Q$-exact action 
with that from the standard action in (\ref{mainresult}). 
We see that when $n> \frac{\Delta}{2L}$ and $n<\frac{\Delta -1}{2L}$  the one loop determinant
in (\ref{mainresult2}) precisely agrees with that from the standard action. 
From our discussion in section \ref{sec.BdyConds.1}, we know that  these Kaluza-Klein modes obey
 both normalizable and supersymmetric boundary conditions. Therefore we see explicitly 
 that we can either use the standard action or the $Q$ exact  action to obtain the 
 partition function.  This then is in accordance with the principle of localization that a 
 $Q$-exact deformation should not change the result for the partition function. 
 However if there are Kaluza-Klein modes  $n$ such that  $\frac{\Delta -1}{2L} <n <\frac{\Delta}{2L}$
 their contribution to the one loop determinant for the standard action differs
 from that of the $Q$-exact action. It is for these modes, that normalizable 
 boundary conditions do not agree with supersymmetric boundary conditions. 
 Therefore, there is no reason why the two one loop determinants, (\ref{mainresult}) and (\ref{mainresult2}), obtained from  the Green's function method using normalizable boundary conditions should agree. \\
Furtermore, it is easy to see that the partition function (\ref{mainresult2.2}) is invariant under $\Delta \rightarrow 2L+\Delta$ for non zero $q$ and arbitrary $L$. Thus this symmetry also continues to hold for $Q$-exact deformations.

\section{One loop determinant from the index of $D_{10}$} \label{indexd10}

As emphasised several times in the text, to define the path integral one should 
use normalizable boundary conditions for all the fields involved. We also saw in section 
\ref{sec.BdyConds.1} that  if there exists an integer $n$ 
in the domain $D$ supersymmetic boundary conditions are not compatible with 
normalizable boundary conditions. For such a situation we expect that  one certainly cannot 
use the localization methods which rely on supersymmetric boundary conditions to obtain 
one loop determinant. 
For the case of supersymmetic boundary conditions one method of obtaining the one loop determinant is to evaluate the 
index of the $D_{10}$ operator associated with the $Q$-exact action in (\ref{Sec:Q-exact.1}).
In this section we do the index computation using the explicit kernel and co-kernel analysis for the $D_{10}$ operator.  We will then compare the result  with the Green's function answer
obtained in (\ref{mainresult2}) and show that when there exists $n$ in the domain $D$ the answers do not agree. 

We start with the $Q-$exact deformation \eqref{eq:Q-exactV.1}, which is given as
\be
V=\int d^{3}x\sqrt{g}\frac{1}{\cosh r}\Big[\bar F\,(\psi\wt\xi)-\psi\g^{\m}\xi\,D_{\m}\bar\phi-iq\sigma\bar\phi\,(\psi\xi)- F\,(\wt\psi\xi)+\wt\psi\g^{\m}\wt\xi\,D_{\m}\phi+iq\sigma\phi\,(\wt\psi\wt\xi)\Big]\,.
\ee
To recover the terms relevant for $D_{10}$ operator, we express the above action as follows
\begin{align}
 V=\text{Tr}\,(Q X_0~ & X_1)\left(
\begin{array}{cc}
 \text{D}_{00} & \text{D}_{01} \\
 \text{D}_{10} & \text{D}_{11} \\
\end{array}
\right)\left(
\begin{array}{c}
X_0 \\
 Q X_1 \\
\end{array}
\right)\,.
\end{align}
In the above $X_{0}=\{\phi,\bar\phi\}$ and $X_{1}=\{B,\bar B\}$.
Now the terms relevant for $D_{10}$ operator are
\be
-\frac{\sqrt{g}}{\cosh^{2}r}[(\xi\g^{\m}\xi)\,B\,D_{\m}\bar\phi+(\wt\xi\g^{\m}\wt\xi)\,\wt B\,D_{\m}\phi]\,.
\ee
It is important to note that we get the same $D_{10}$ operator from \eqref{eq:QexactS.1}. \\
Now let us consider first the kernel equation
\be
(\xi\gamma^\mu\xi)D_\mu\bar\phi=0\,.
\ee
This gives the following first order differential equation
\be
\sinh r\,(-n+\frac{\Delta}{2L})\bar f_{n,p}(r)+\frac{p}{L\sinh r}\bar f_{n,p}(r)+\frac{\cosh r}{L}\p_r\bar f_{n,p}(r)=0\,.
\ee
In the above we have expanded the field in terms of Fourier modes $\bar\phi=\sum_{n,p}e^{-i(n\tau+p\theta)}\bar f_{n,p}(r)$. The solution of the above differential equations is given by
\be\label{eq:Soln.fb.1}
\bar f_{n,p}(r)=C_1(\cosh r)^{\frac{1}{4}(2Ln+4p-\Delta)}(\sinh r)^{\frac{1}{4}(-2Ln-4p+\Delta)}(\sinh 2r)^{-\frac{1}{4}(-2Ln+\Delta)}\,.
\ee
We see that as $r\rightarrow \infty$, the solution goes like $\sim e^{-\frac{r}{2}(-2Ln+\Delta)}$ and for $r\rightarrow 0$, the solution goes like $\sim r^{-p}$. Clearly the solution with $p>0$ is not smooth near $r\rightarrow 0$. Therefore, the solutions with $p>0$ are excluded.  \\
The other kernel equation is
\be
(\wt\xi\gamma^\mu\wt\xi)D_\mu\phi=0\,,
\ee
which becomes
\be
\sinh r\,(-n+\frac{\Delta}{2L})f_{n,p}(r)+\frac{p}{L\sinh r}f_{n,p}(r)+\frac{\cosh r}{L}\p_rf_{n,p}(r)=0\,.
\ee
In the above we have expanded the field in terms of Fourier modes $\phi=\sum_{n,p}e^{i(n\tau+p\theta)} f_{n,p}(r)$.
It is again the same equation as above and, therefore it has the same solution.\\
Now we look for the cokernel equations. In this case we get
\be
D_\mu\left(\frac{\xi\gamma^\mu\xi}{\cosh^{2}r}B\right)=0\,.
\ee
Using the Killing spinor equation we get
\be
\sinh r\,(n-\frac{\Delta}{2L})b_{n,p}(r)+\frac{1-p}{L\sinh r}b_{n,p}(r)+\frac{\cosh r}{L}\p_rb_{n,p}(r)=0\,.
\ee
In the above we have expanded the field in terms of Fourier modes $B=\sum_{n,p}e^{i(n\tau+(p-1)\theta)} b_{n,p}(r)$. We see that it is exactly same as the above equations (equation for $\bar\phi$). Therefore the solution is
\be\label{eq:Soln.b.1}
b_{n,p}(r)=C_2(\cosh r)^{\frac{1}{4}(-2Ln-4p+4+\Delta)}(\sinh r)^{\frac{1}{4}(2Ln+4p-4-\Delta)}(\sinh 2r)^{-\frac{1}{4}(2Ln-\Delta)}\,.
\ee
The asymptotic of the above solution is
\be
b_{n,p}\sim e^{-\frac{r}{2}(2Ln-\Delta)}\,\,\, \text{for}\,\,r\rightarrow\infty,\,\,\text{and}\,\,\, b_{n,p}\sim r^{p-1}, \,\, r\rightarrow 0\,.
\ee
Clearly for $p<1$ the solution is not smooth and therefore excluded.  Similarly we have another cokernel equation
\be
D_\mu\left(\frac{\wt\xi\gamma^\mu\wt\xi}{\cosh^{2}r}\wt B\right)=0\,,
\ee
and using the killing spinor equation we get
\be
\sinh r\,(n-\frac{\Delta}{2L})\wt b_{n,p}(r)+\frac{1-p}{L\sinh r}\wt b_{n,p}(r)+\frac{\cosh r}{L}\p_r\wt b_{n,p}(r)=0\,.
\ee
In the above we have expanded the field in terms of Fourier modes $\wt B=\sum_{n,p}e^{-i(n\tau+(p-1)\theta)} \wt b_{n,p}(r)$. We see that it is exactly same as the above equation and therefore, the solution remains same.
\subsection{ Evaluation of the index}

We now impose the following susy boundary conditions at asymptotic infinity and 
evaluate the index of $D_{10}$ operator. 
The fall off conditions at infinity are 
\be\label{eq:susyBdy.1}
e^{r/2}f_{n,p}\rightarrow 0,\quad e^{-r/2}b_{n,p}\rightarrow 0\,.
\ee
Similar boundary conditions are imposed on the complex conjugate fields.
In this case we see from the asymptotic behaviour of the solutions \eqref{eq:Soln.fb.1} and \eqref{eq:Soln.b.1} that the mode $\bar f_{n,p}$ belongs to the kernel if $n$ satisfies $-2Ln+\Delta-1>0$ and the mode $b_{n,p}$ belongs to co-kernel if $n$ satisfies $2Ln-\Delta+1>0$. Thus the contribution to index from fields $(\phi,\bar\phi)$ and $(B,\bar B)$ are
\bea
&& \phi : \prod^{-\infty}_{n<\frac{\Delta-1}{2L},p=0}\left[i(n+\frac{p}{L})+iq\Lambda-i\frac{\Delta}{2L}\right]\,,\\
&& \bar\phi : \prod^{-\infty}_{n<\frac{\Delta-1}{2L},p=0}\left[-i(n+\frac{p}{L})-iq\Lambda+i\frac{\Delta}{2L}\right]\,,\\
&& B : \prod^{\infty}_{n>\frac{\Delta-1}{2L},p=1}\left[i(n+\frac{p-1}{L})+iq\Lambda-\frac{i}{2L}(\Delta-2)\right]\,,\\
&& \bar B : \prod^{\infty}_{n>\frac{\Delta-1}{2L},p=1}\left[-i(n+\frac{p-1}{L})-iq\Lambda+\frac{i}{2L}(\Delta-2)\right]\,.
\eea
Here $\Lambda=i\a$. This implies the one loop partition function is given by\footnote{If we relax the strict normalizable boundary condition on $f_{n,p}$ \eqref{eq:susyBdy.1} and allow $e^{r/2}f_{n,p}\sim \mathcal O(1)$ as $r\rightarrow \infty$, then the corresponding partition function is $Z^{{\rm index}}=\frac{ \prod^{\infty}_{n>\frac{\Delta-1}{2L},p=0}\left[i(n+\frac{p}{L})+iq\Lambda-\frac{i}{2L}(\Delta-2)\right]\,}{\prod^{\infty}_{n\geq\ceil{\frac{1-\Delta}{2L}},p=0}\left[i(n+\frac{p}{L})-iq\Lambda+i\frac{\Delta}{2L}\right]\,}\,$. This agrees with the Green's function answer if there are no $n$'s in $D$.}
\be
Z^{{\rm index}}=\frac{ \prod^{\infty}_{n>\frac{\Delta-1}{2L},p=0}\left[i(n+\frac{p}{L})+iq\Lambda-\frac{i}{2L}(\Delta-2)\right]\,}{\prod^{\infty}_{n>\frac{1-\Delta}{2L},p=0}\left[i(n+\frac{p}{L})-iq\Lambda+i\frac{\Delta}{2L}\right]\,}\,.
\ee
We rewrite this in terms the free energy, we will also remove the factors of $i$ and multiply 
$L$ in the numerator and denominator.  This of course involves scaling the 
one loop determinant with an infinite $L$ dependent constant. 
We obtain
\begin{eqnarray}\label{mainresult3}
\ln Z^{ {\rm index} } = \sum_{p=1, n> \frac{\Delta-1}{2L}} 
\ln ( p + L ( n + i q \alpha) - \frac{\Delta}{2} )  - \sum_{p=0, n > \frac{1-\Delta}{2L}}
\ln ( p + L ( n -i q\alpha) + \frac{\Delta}{2})\,.  \nonumber \\
\end{eqnarray}
Let us   compare the results for the one loop determinant for the standard 
action in (\ref{mainresult}) and for the $Q$-exact action in (\ref{mainresult2}) 
using the Green's function obeying normalizable boundary conditions 
with (\ref{mainresult3}). We see that all the three results for the one loop determinant 
agree when there exists no integer $n$ in the domain $D$ and $\frac{\Delta-1}{2L}$ is not an integer. 
From our discussion on boundary conditions we see that this is expected since 
for this situation the fields obeying normalizable boundary conditions 
also satisfy supersymmetric boundary conditions. 

\noindent
{\bf{Discontinuity in the index:}}
We have seen that the one loop determinant of the standard action 
at $L=1$ has the symmetry $\Delta \rightarrow 2-\Delta$. 
This implies that we can define it  such that it is continuous at $\Delta =1$. 
Let us examine the behaviour of the index in (\ref{mainresult3}) for $L=1$. 
When  $0\leq\Delta<1, L=1$, the result for  the index is given by
\be
Z^{{\rm index}}_{ 0\leq\Delta<1, L=1}
 =\prod_{m=1}^\infty\left(\frac{m+q\Lambda-\frac{\Delta}{2}}{m-q\Lambda+\frac{\Delta}{2}}\right)^m\,.
\ee
Taking the limit $\Delta \rightarrow 1^{-}$ we obtain
\begin{equation}\label{indless}
Z_{<} = Z^{{\rm index}}|_{\Delta \rightarrow 1^{-}, L=1}=\prod_{m=1}^\infty\left(\frac{m+q\Lambda-\frac{1}{2}}{m-q\Lambda+\frac{1}{2}}\right)^m\,.
\end{equation}
Similarly, for the case $L=1$ and $1<\Delta<2$, the result of the one loop determinant is given by
\be
 Z^{{\rm index}}_{1<\Delta<2, L=1 }
=\prod_{m=1}^\infty\left(\frac{m+q\Lambda-\frac{\Delta-2}{2}}{m-q\Lambda+\frac{\Delta-2}{2}}\right)^m\,.
\ee
Lets now take the limit $\Delta \rightarrow 1^{+} $ to get 
\be\label{indgreat}
Z_{>}= Z^{{\rm index}}_{\Delta\rightarrow 1^{+}, L=1}
=\prod_{m=1}^\infty\left(\frac{m+q\Lambda+\frac{1}{2}}{m-q\Lambda-\frac{1}{2}}\right)^m\,.
\ee
Taking the ratio of (\ref{indless}) and (\ref{indgreat}) we find 
\begin{equation}
\frac{Z_{<} }{Z_{>}} = \prod_{n=0}^\infty \left[ ( n+\frac{1}{2})^2 -  (q\Lambda)^2 \right]\,.
\end{equation}
Thus there is a jump in the index at $\Delta =1$\,. It is important to note that this feature of the supersymmetric index is not consistent with the partition function of the free theory presented in appendix \ref{appendixb} where it is continuous in $\Delta$\,.

\section{Conclusions} \label{conclusions}

We have developed the Green's function method to obtain the one loop determinant 
of the chiral multiplet  on $AdS_2\times S^1$. It is 
coupled to a  background vector multiplet which preserves supersymmetry. 
We implemented normalizable boundary conditions on the   Green's function  in $AdS_2$\,. 
Our study in this example shows that when fields, which satisfy  normalizable boundary conditions,
do not obey supersymmetric boundary conditions the partition function 
does depend on the  $Q$-exact deformation. 
Furthermore, the one loop determinants evaluated using 
the index of the $D_{10}$ operator associated with the 
localizing action do not agree with that using normalizable boundary conditions when 
these conditions are such that they are not consistent with 
supersymmetric boundary conditions. This is because the action of $Q$ is non normalizable {\it{i.e.}} it takes the space of normalizable wave functions to a space which includes non normalizable wave functions.
Therefore,  care should be taken when one applies the method of localization to evaluate
supersymmetric observables for field theories on
non-compact spaces\,. \\
When the normalizable and supersymmetric boundary conditions are compatible, the result using the Green's function method
localizes at the fixed point (which is the origin of $AdS_{2}$) and the boundary, and agrees with the result obtained from index calculation by evaluating Kernel and CoKernel
of $D_{10}$ operator. This feature is certainly a reflection of  supersymmetry
and we hope to prove this in general. As the result in our approach is localized at the fixed point and the boundary, it can be calculated more easily by analysing
Green's functions locally and does not require finding global solutions of $D_{10}$ operator. The latter in more complicated cases, such as vector multiplets 
and higher dimensional spaces are highly coupled differential equations, and global solutions to them are very difficult to construct. \\
One surprising  property we noted in our study of the one loop determinant  of the chiral multiplet
on $AdS_2\times S^1$ is  when 
the radius of $AdS_2$ equals that of $S^1$  there exists a hidden symmetry\footnote{We noted in the main text that there is also a symmetry $\Delta \rightarrow 2+\Delta$ when $q$ is non zero.} 
$\Delta \rightarrow 2-\Delta$, where $\Delta$ is the R-charge of the chiral multiplet. 
This symmetry is manifest in the eigen function
 representation of the partitions function when the chiral multiplet is decoupled from the background vector multiplet\,. We have seen that it persists 
 when the coupling is turned on. 
 One implication of this symmetry is the following: Consider the case 
 when $\Delta = \frac{1}{2}$. This is the situation when the  theory is also conformal. 
 For this situation there always exists  one integer namely $n=0$ for which 
 the  Kaluza-Klein modes  which  obeys normalizable boundary conditions are not
 compatible with supersymmetric boundary conditions. However due to the hidden symmetry,
 the result for the one loop determinant can be obtained
 by examining the situation at $\Delta = \frac{3}{2}$ for which the boundary conditions 
 are always compatible with supersymmetry. 
 This is surprising and perhaps this feature is useful. 
  Therefore it is important to understand if this property persists more generally 
  when the vector multiplet  is  no longer just a classical background.  Such properties 
  if true in general will be useful in the application of localization on such backgrounds.

Localization has been used to evaluate partition functions on $AdS_2\times S^2$ with 
supersymmetric backgrounds corresponding to extremal black holes \cite{Dabholkar:2010uh,Dabholkar:2011ec,Gupta:2012cy,Dabholkar:2014ema,Murthy:2015yfa,Gupta:2015gga}.  
In light of our results on $AdS_2\times S^1$, we have re-examined this question for the 
partition function of the  hypermutliplet on $AdS_2\times S^2$ \cite{David:2018}. We find that the normalizable  boundary conditions for fields in extremal black hole backgrounds on $AdS_2\times S^2$ always satisfy 
supersymmetric  boundary conditions. 
We can also ask the same questions for  the evaluation of  partition function 
of the vector multiplets on both $AdS_2\times S^1$ as well as that of
$AdS_2\times S^2$ using localization.
We hope to report on these questions  in the near future.

\acknowledgments
We thank Sameer Murthy and Alejandro Cabo Bizet for useful conversations. We would especially like to thank Ashoke Sen for useful discussions on boundary conditions in defining path integrals. The work of RG is supported by the ERC Consolidator Grant N. 681908, ``Quantum black holes:
A macroscopic window into the microstructure of gravity".

\appendix 
\section{Supersymmetry on AdS$_{2}\times$S$^{1}$}
\label{appendixa}

\subsection{Conventions}
The covariant derivative of a fermion is given by
\be
\nabla_\mu\psi=\left(\p_\mu+\frac{i}{4}\omega_{\mu\,ab}\varepsilon^{abc}\gamma_c\right)\psi,\qquad \varepsilon^{123}=1.
\ee
Our choice of gamma matrices are
\be
\gamma^1=\begin{pmatrix}1&0\\0& -1\end{pmatrix},\quad \gamma^2=\begin{pmatrix}0&-1\\-1& 0\end{pmatrix},\quad \gamma^3=\begin{pmatrix}0&i\\-i& 0\end{pmatrix}\,.
\ee
They satisfy gamma matrices algebra
\be
\gamma^a\gamma^b=\delta^{ab}+i\varepsilon^{abc}\gamma_c\,.
\ee
\be
\gamma^{aT}=-C\gamma^aC^{-1},\quad C=\begin{pmatrix}0 & 1\\-1 & 0\end{pmatrix},\quad C^T=-C=C^{-1}\,.
\ee
In Lorentzian space $\psi$ and $\bar\psi$ are complex conjugate to each other but in Euclidean space fermions $\psi$ and $\bar\psi$ are independent two component complex spinor. The product of two fermions $\epsilon$ and $\psi$ is defined through charge conjugation matrix
\be
\epsilon\psi=\epsilon^TC\psi\,.
\ee

\subsection{Killing spinors}\label{Killingspinor}
The Killing spinor equations are given by
\bea\label{}
\left(\nabla_\mu-iA_\mu\right)\epsilon=-\frac{1}{2}H\gamma_\mu\epsilon-i V_\mu\epsilon-\frac{1}{2}\epsilon_{\mu\nu\rho}V^\nu\gamma^\rho\epsilon\,,\nn\\
\left(\nabla_\mu+iA_\mu\right)\tilde\epsilon=-\frac{1}{2}H\gamma_\mu\tilde\epsilon+i V_\mu\tilde\epsilon+\frac{1}{2}\epsilon_{\mu\nu\rho}V^\nu\gamma^\rho\tilde\epsilon\,.
\eea
Here $\varepsilon^{\mu\nu\rho}=\frac{1}{\sqrt{g}}\epsilon^{\mu\nu\rho},\quad \epsilon^{\tau\eta\theta}=1$.\\The background metric  of $AdS_2 \times S^1$ is given by
\be\label{metric}
ds^2=d\tau^2+L^2(dr^2+\sinh^2r \,d\theta^2)\,.
\ee
The vielbein are $e^1=d\tau,\, e^2=L \,dr,\, e^3=L\sinh r \,d\theta$.\\ 
The non vanishing components of Christoffel symbols and spin connections are given by
\be
\Gamma^r_{\theta\theta}=-\cosh r\sinh r,\quad \Gamma^\theta_{r\theta}=\coth r,\quad \omega^3{}_2=\cosh r\, d\theta\,.
\ee
The solution of killing spinor equations is given as
\bea\label{susy-backgrd}
&&\epsilon= e^{\frac{i\theta}{2}}\begin{pmatrix}i\cosh(\frac{r}{2})\\\sinh(\frac{r}{2})\end{pmatrix}\,,\qquad \tilde\epsilon=e^{-\frac{i\theta}{2}}\begin{pmatrix}\sinh(\frac{r}{2})\\i\cosh(\frac{r}{2})\end{pmatrix}\,,\nn\\ &&A_\tau=V_\tau=\frac{1}{L}\,,\quad A_{r,\theta}=H=0\,.
\eea

\subsection{Localization for Vector Multiplet on AdS$_{2}\times$S$^{1}$}
\be
\Psi=\frac{i}{2}(\tilde\epsilon\lambda+\epsilon\tilde\lambda)\,,\quad \Psi_\mu=Qa_\mu=\frac{1}{2}(\epsilon\gamma_\mu\tilde\lambda+\tilde\epsilon\gamma_\mu\lambda)\,.
\ee
The fermion bilinears are convenient for the index computation. The inverse of the above relations express $(\lambda,\tilde\lambda)$ in terms of  $\Psi,\Psi_\mu$ are given by
\bea
\lambda=\frac{1}{\tilde\epsilon\epsilon}\left[\gamma^\mu\epsilon\Psi_\mu-i\epsilon\Psi\right],\quad \tilde\lambda=\frac{1}{\epsilon\tilde\epsilon}\left[\gamma^\mu\tilde\epsilon\Psi_\mu-i\tilde\epsilon\Psi\right]\,.
\eea
The supersymmetry transformation of the bilinears are
\bea
&&Q\Psi=\frac{1}{4}(\tilde\epsilon\epsilon)G-\frac{i}{2}\l(\tilde\epsilon\gamma^{\mu\nu}\epsilon\r) F_{\mu\nu}-\frac{1}{L}\sigma\,,\nn\\
&&Q\Psi_\mu=\mathcal L_K a_\mu+D_\mu\Lambda\,.
\eea
Here $\Lambda=\tilde\epsilon\epsilon\,\sigma-K^\rho a_\rho$. Next we deform the action by a $Q$-exact term, $t\,QV_{loc}$. According to the principle of supersymmetric localization, the partition function does not depend on the parameter $t$ and the choice of $V_{loc}$. Thus one can take $t$ to infinity. In this limit the path integral receives contribution from the field configurations which are minima of $QV_{loc}$. 
One convenient choice of $V_{loc}$ is given by
\be
V_{loc}=\int d^3x\sqrt{g}\frac{1}{(\tilde\epsilon\epsilon)^2}\text{Tr}\left[\Psi^\mu (Q\Psi_\mu)^\dagger+\Psi (Q\Psi)^\dagger\right]\,.
\ee
The bosonic part of the $QV_{loc}$ action is given by
\bea\label{bosonLagr}
QV_{loc\{\text{bosonic}\}}&=&\int d^3x\sqrt{g}\frac{1}{2(\tilde\epsilon\epsilon)^2}\text{Tr}\left[(Q\Psi^\mu)(Q\Psi_\mu)^\dagger+(Q\Psi)(Q\Psi)^\dagger\right]\nn\\&=&\int d^3x\sqrt{g}\text{Tr}\left[\frac{1}{4}F_{\mu\nu}F^{\mu\nu}-\frac{1}{2\cosh^2r}D_\mu(\cosh r\,\sigma)D^\mu(\cosh r\,\sigma)-\frac{1}{32}\left(G-\frac{4\sigma}{L\cosh r}\right)^2\right]\,.\nn\\
\eea
The minima of  $QV_{loc\{\text{bosonic}\}}$ are the solutions of the following equations
\be
F_{\mu\nu}=0\,,\quad D_\mu(\cosh r\,\sigma)=0\,,\quad G=\frac{4\sigma}{L\cosh r}\,.
\ee 
Thus the solution of localization equation upto gauge transformations is given by
\be
a_\mu=0\,,\quad\sigma=\frac{i\alpha}{\cosh r}\,,\quad G=\frac{4i\alpha}{L\cosh^2r}\,.
\ee
Here $\alpha$ is a real constant matrix valued in Lie algebra.
\section{Eigen function method for the free chiral multiplet} \label{appendixb}

In this section we will compute the partition function of a free chiral multiplet on AdS$_{2}\times$S$^{1}$ using the periodic boundary conditions along the $S^{1}$ direction for both the scalar and fermion. The free chiral multiplet is one loop exact and therefore, the one loop determinant can be evaluated exactly. In this section we compute the one loop determinant by expanding the fields in terms of harmonics on AdS$_{2}$. 
We follow \cite{Klebanov:2011uf} where the eigen function method 
was used to obtain the free energy of a conformal  scalar and massless fermions 
on $AdS_2 \times S^1$. In \cite{Klebanov:2011uf} the fermions obeyed thermal  boundary 
conditions  on the $S^1$,  here we impose periodic boundary conditions 
for the fermions. 
We can then compare the result with the explicit Green's function calculation presented in the section \ref{1loopst}. \\The Lagrangian for free chiral multiplet is given as
\be
\mathcal L=\mathcal D^\mu\bar\phi \mathcal D_\mu\phi-\bar\psi\gamma^\mu \mathcal D_\mu\psi-\bar FF-\frac{\Delta}{4}R\,\bar\phi\phi+\frac{1}{2}(\Delta-\frac{1}{2})V^2\bar\phi\phi\,.
\ee
Here $\Delta$ is the R-charge of the chiral multiplet $i.e.$ the R-charges of fields $(\phi,\psi, F)$ are $(\Delta,\Delta-1,\Delta-2)$ and the covariant derivative 
\be
\mathcal D_\mu=\nabla_\mu-i\Delta_{R}(A_\mu-\frac{3}{2}V_\mu)-i(\Delta_{R}-r_{0})V_\mu\,,
\ee
where $\Delta_{R}$ is the R-charge of the field and $r_{0}=1/2$ for scalar and $r_{0}=-1/2$ for fermion.\\ 
After integrating out the auxiliary field and substituting $R=2V^{2}$, we get
\be
\mathcal L=\mathcal D^\mu\bar\phi \mathcal D_\mu\phi-\bar\psi\gamma^\mu \mathcal D_\mu\psi-\frac{1}{4}V^{2}\,\bar\phi\phi\,.
\ee
Now let us consider first the scalar terms in the Lagrangian which are 
\be
\mathcal L^s=\Big(\nabla_\mu+\frac{i}{2}(\Delta-1)V_\mu\Big)\bar\phi \,\Big(\nabla_\mu-\frac{i}{2}(\Delta-1)V_\mu\Big)\phi-\frac{1}{4}V^{2}\bar\phi\phi\,.
\ee
Now we define $\phi=e^{\frac{i}{2L}(\Delta-1)\tau}\eta$. Then the Lagrangian becomes
\be
\mathcal L^s=\nabla_\mu\bar\eta \nabla_\mu\eta-\frac{1}{4L^{2}}\bar\eta\eta\,.
\ee
Now expanding each KK mode, labelled by an integer $n$, in terms of scalar harmonics on AdS$_{2}$ with eigen value $(\l^{2}+\frac{1}{4})/L^{2}$ with the density of states \cite{Camporesi:1994ga}
\be
\mathcal D(\lambda)\,d\lambda=-\l\,\tanh(\pi\lambda)\,d\lambda\,,
\ee
the one loop Free energy $F=-\ln Z$ is given by
\be
F^s=\sum_{n\in\mathbb Z}\int_0^\infty d\lambda \mathcal D(\lambda)\ln\Big(\frac{\lambda^{2}}{L^{2}}+\Big(n-\frac{\Delta-1}{2L}\Big)^2\Big)\,.
\ee
Now using the regularized sum (see the appendix B of \cite{Klebanov:2011uf})
\be
\sum_{n\in\mathbb Z}\ln(a^{2}+\frac{(n+\a)2}{q^2})=\ln\Big[2\cosh(2\pi q|a|)-2\cos(2\pi\a)\Big]\,,
\ee
we obtain the contribution to the free energy from the scalar field 
\be
F^s=\frac{Vol(H_2)}{2\pi}\int_0^\infty\,d\lambda\,\lambda\,\tanh(\pi\lambda)\,\ln[2\cosh(\frac{2\pi\lambda}{L})-2\cos(\frac{\pi}{L}(1-\Delta))]\,.
\ee
Now we look at the fermion. We have the following Lagrangian
\be
\mathcal L^f=-\bar\psi\gamma^\mu \mathcal D_\mu\psi=-\bar\psi\gamma^\mu(\nabla_\mu-\frac{i\Delta}{2}V_\mu)\psi\,.
\ee
Now we define $\psi=e^{\frac{i\Delta}{2L}\tau}\theta$, then we get the fermionic Lagrangian
\be
\mathcal L^f=-\bar\th\gamma^\mu \nabla_\mu\th\,.
\ee
As in the scalar case we expand each KK mode in terms of harmonics of a Dirac operator on AdS$_{2}$ labelled by eigen value $\pm i\frac{\l}{L}$ with density of states
\be
\tilde{\mathcal D} (\lambda)\,d\lambda=-2\lambda\coth(\pi\lambda)\,d\lambda\,.
\ee
Then the free energy of the periodic fermionic field is given by
\be
F^f=-\frac{1}{2}\sum_{n\in\mathbb Z}\int_0^\infty\,d\lambda\,\tilde{\mathcal D} (\lambda)\ln[\frac{\lambda^2}{L^{2}}+(n-\frac{\Delta}{2L})^2]=-\frac{1}{2}\int_0^\infty\,d\lambda\,\tilde{\mathcal D} (\lambda)\ln[2\cosh(2\pi\frac{\lambda}{L})-2\cos(\frac{\pi\Delta}{L})]\,.
\ee
Thus the complete free energy is
\bea \label{oneloopint}
F=F^s+F^f&=&-\int\,d\lambda\,\lambda\,\tanh(\pi\lambda)\ln\left[2\cosh(2\pi\frac{\lambda}{L})-2\cos(2\pi\frac{\Delta-1}{2L})\right]\nn\\ &&+\int\,d\lambda\,\lambda\,\coth(\pi\lambda)\ln\left[2\cosh(2\pi\frac{\lambda}{L})-2\cos(2\pi\frac{\Delta}{2L})\right]\,.
\eea
Thus to get the free energy and the partition function we need to perform the above one dimensional integral. For example in the case of $L=1$ with $\Delta=\frac{1}{2}$ 
\footnote{In this case, the scalar theory is  conformal on $AdS_2$}, we get
\be
F_{\Delta=\frac{1}{2}}=\frac{1}{4\pi}(\frac{\pi}{2}\ln 2+2\, \text{Catalan}),\quad \text{and}\,\,\,\ln Z_{\Delta=\frac{1}{2}}=-\frac{1}{4\pi}(\frac{\pi}{2}\ln 2+2\, \text{Catalan})\,,
\ee
and with $\Delta=0$, we get
\be
F_{\Delta=0}=0\,, \quad \text{and}\,\,\, Z_{\Delta=0}=1\,.
\ee
Also note that for $\Delta=1$ the scalar $\eta$ has periodic boundary condition and fermion $\theta$ has anti periodic boundary condition along S$^{1}$. In this case we get
\be
F_{\Delta=1}=\frac{1}{2}\ln 2\,.
\ee

Below we will present the result for the free energy for general $\Delta$ and, $L=1$ and $L=2$.\\

\noindent
{\bf Case 1: \bf{Result for $L=1$ and $0\leq\Delta<2$}}\\
We will now present the result for general $\Delta$ lying in the range $[0,2)$. We begin with the expression
\bea
F=-\int_0^\infty \,d\lambda\,\lambda\,\Big[\tanh(\pi\lambda)\,\ln\left[2\cosh(2\pi\lambda)+2\cos(\pi\Delta)\right]\nn\\-\coth(\pi\lambda)\,\ln\left[2\cosh(2\pi\lambda)-2\cos(\pi\Delta)\right]\Big]\,.
\eea
We notice that the integrand has the symmetry $\Delta\rightarrow 2-\Delta$. Now, we calculate its first derivative with respect to $\Delta$
\bea
\frac{dF}{d\Delta}=\pi\sin(\pi\Delta)\int_0^\infty \,d\lambda\,\lambda\,\Big[\tanh(\pi\lambda)\,\frac{1}{\cosh(2\pi\lambda)+\cos(\pi\Delta)}\nn\\+\coth(\pi\lambda)\,\frac{1}{\cosh(2\pi\lambda)-\cos(\pi\Delta)}\Big]\,.
\eea
Now the integrals on the RHS is calculable which are given as 
\bea
\int_0^\infty \,d\lambda\,\lambda\,\tanh(\pi\lambda)\,\frac{1}{\cosh(2\pi\lambda)+\cos(\pi\Delta)}=\frac{1}{48\pi^{2}\sin^{2}(\frac{\pi\Delta}{2})}\Big[\pi^{2}+6Li_{2}(-e^{i\pi\Delta})\nn\\+6Li_{2}(-e^{-i\pi\Delta})\Big]\,,
\eea
and
\bea
\int_0^\infty \,d\lambda\,\lambda\,\coth(\pi\lambda)\,\frac{1}{\cosh(2\pi\lambda)-\cos(\pi\Delta)}=-\frac{1}{24\pi^{2}\sin^{2}(\frac{\pi\Delta}{2})}\Big[-\pi^{2}+3Li_{2}(e^{i\pi\Delta})\nn\\+3Li_{2}(e^{-i\pi\Delta})\Big]\,.
\eea
Thus we get the following first order differential equation
\be
\frac{dF}{d\Delta}=-\frac{\sin(\pi\Delta)}{16\pi\sin^{2}(\frac{\pi\Delta}{2})}\Big[-\pi^{2}+2Li_{2}(e^{i\pi\Delta})+2Li_{2}(e^{-i\pi\Delta})-2Li_{2}(-e^{i\pi\Delta})-2Li_{2}(-e^{-i\pi\Delta})\Big]\,.
\ee
{\bf Case : $0\leq\Delta<1$}\\
Now we use the following identity to simplify the expression of the dilogaritm 
\be
Li_{2}(e^{2i\pi x})+Li_{2}(e^{-2i\pi x})=2\pi^{2}(x^{2}-x+\frac{1}{6})\,\qquad \text{for}\,\,0\leq\text{Re}\,x<1\,.
\ee
In this case we get
\bea\label{eq:AdS2;Free,diff,R<1}
\frac{dF}{d\Delta}&=&\frac{\sin(\pi\Delta)}{8\pi(\cos(\pi\Delta)-1)}\Big[-\pi^{2}+2Li_{2}(e^{i\pi\Delta})+2Li_{2}(e^{-i\pi\Delta})-2Li_{2}(e^{i\pi(\Delta+1)})-2Li_{2}(e^{-i\pi(\Delta+1)})\Big]\,,\nn\\&=&\frac{\sin(\pi\Delta)}{8\pi(\cos(\pi\Delta)-1)}\Big[-\pi^{2}+4\pi^{2}\left(\frac{\Delta^{2}}{4}-\frac{\Delta}{2}+\frac{1}{6}\right)-4\pi^{2}\left(\frac{(\Delta+1)^{2}}{4}-\frac{\Delta+1}{2}+\frac{1}{6}\right)\Big]\,,\nn\\&=&\frac{\pi\Delta}{4}\cot\Big(\frac{\pi\Delta}{2}\Big)\,.
\eea
Now integrating over $\Delta$ we get
\be
F=\frac{\Delta}{2}\ln(1-e^{i\pi\Delta})-\frac{i}{2\pi}\Big(\frac{\pi^{2}\Delta^{2}}{4}+Li_{2}(e^{i\pi\Delta})\Big)+C_{1}\,,
\ee
where $C_{1}$ is some integration constant which we determine by requiring that $F|_{\Delta=0}=0$. 
\be
-\frac{i}{2\pi}Li_{2}(1)+C_{1}=0\,\Rightarrow C_{1}=\frac{i}{2\pi}\zeta(2)\,.
\ee
Thus the free energy in this case is
\be\label{eq:AdS2;Free;R<1}
F=\frac{\Delta}{2}\ln(1-e^{i\pi\Delta})-\frac{i}{2\pi}\Big(\frac{\pi^{2}\Delta^{2}}{4}+Li_{2}(e^{i\pi\Delta})-\zeta(2)\Big)\,.
\ee
{\bf Case: $1\leq \Delta<2$}\\
In this case we get
\bea\label{eq:AdS2;Free,diff,R<2}
\frac{dF}{d\Delta}&=&\frac{\sin(\pi\Delta)}{8\pi(\cos(\pi\Delta)-1)}\Big[-\pi^{2}+2Li_{2}(e^{i\pi\Delta})+2Li_{2}(e^{-i\pi\Delta})-2Li_{2}(e^{i\pi(\Delta-1)})-2Li_{2}(e^{-i\pi(\Delta-1)})\Big]\,,\nn\\&=&\frac{\sin(\pi\Delta)}{8\pi(\cos(\pi\Delta)-1)}\Big[-\pi^{2}+4\pi^{2}\left(\frac{\Delta^{2}}{4}-\frac{\Delta}{2}+\frac{1}{6}\right)-4\pi^{2}\left(\frac{(\Delta-1)^{2}}{4}-\frac{\Delta-1}{2}+\frac{1}{6}\right)\Big]\,,\nn\\&=&\frac{\pi(2-\Delta)}{4}\cot\Big(\frac{\pi\Delta}{2}\Big)\,.
\eea
Integrating with respect to $\Delta$ we get
\be\label{eq:AdS2;Free;R<2}
F=\frac{2-\Delta}{2}\ln(1-e^{i\pi(2-\Delta)})-\frac{i}{2\pi}\Big(\frac{\pi^{2}(2-\Delta)^{2}}{4}+Li_{2}(e^{i\pi(2-\Delta)})-\zeta(2)\Big)\,.
\ee
{\bf Case 2: \bf{Result for $L=2$ and $0\leq\Delta<2$}}\\
In this case we get
\bea
\frac{dF^{s}}{d\Delta}=-\frac{\pi}{2}\sin\Big(\frac{\pi}{2}(\Delta-1)\Big)\int^{\infty}_{0}\,d\lambda\,\l\tanh(\pi\lambda)\frac{1}{\cosh\pi\lambda-\cos\Big(\frac{\pi}{2}(\Delta-1)\Big)}\,.
\eea
Integrating RHS we get
\be
\frac{dF^{s}}{d\Delta}=\frac{1}{48\pi}\cot(\frac{\pi\Delta}{2})\Big[\pi^{2}+24Li_{2}(-ie^{\frac{i\pi\Delta}{2}})+24Li_{2}(ie^{-\frac{i\pi\Delta}{2}})\Big]\,.
\ee
Now we evaluate the above expression for different ranges of R-charge $\Delta$. For the case when ${0\leq\Delta<1}$, we get
\be
\frac{dF^{s}}{d\Delta}=\frac{\pi}{16}\Big[3-(1-\Delta)(\Delta+3)\Big]\cot\frac{\pi\Delta}{2}\,.
\ee
For the case when ${1\leq\Delta<2}$, we get
\be
\frac{dF^{s}}{d\Delta}=\frac{\pi}{16}\Big[3-(\Delta-1)(5-\Delta)\Big]\cot\frac{\pi\Delta}{2}\,.
\ee
In the case of fermion we get for ${0\leq\Delta<2}$
\be\label{eq:AdS2.2;Free;R<1}
\frac{dF^{f}}{d\Delta}=-\frac{\pi}{8\sin\Big(\frac{\pi\Delta}{2}\Big)}\Big[-1+\cos\Big(\frac{\pi\Delta}{2}\Big)+2\cos\Big(\frac{\pi\Delta}{2}\Big)\Big(\frac{\Delta^{2}}{4}-\Delta\Big)\Big]\,.
\ee
Adding the above two for the case when ${0\leq\Delta<1}$, total derivative of free energy is
\be\label{eq:AdS2:L=2.1}
\frac{dF}{d\Delta}=\frac{dF^{f}}{d\Delta}+\frac{dF^{s}}{d\Delta}=\frac{\pi}{8\sin\Big(\frac{\pi\Delta}{2}\Big)}\Big[1+(3\Delta-1)\cos\Big(\frac{\pi\Delta}{2}\Big)\Big]\,.
\ee
Integrating the above we get
\be
F=\frac{1}{16}\Big(-3\pi i\Delta^{2}+12\Delta\ln(1-e^{i\pi\Delta})-4\ln 2-8\ln(\cos\frac{\pi \Delta}{4})\Big)-\frac{3i\pi}{4} Li_{2}(e^{i\pi \Delta})+C_{1}\,,
\ee
where $C_{1}$ is an integration constant. \\
Similarly, for the case when ${1\leq\Delta<2}$ we have
\be\label{eq:AdS2:L=2.2}
\frac{dF}{d\Delta}=\frac{dF^{f}}{d\Delta}+\frac{dF^{s}}{d\Delta}=\frac{\pi}{8\sin\Big(\frac{\pi\Delta}{2}\Big)}\Big[1+(3-\Delta)\cos\Big(\frac{\pi\Delta}{2}\Big)\Big]\,.
\ee
Integrating the RHS, we obtain
\be
F=-\frac{\Delta}{4}\ln(1-e^{i\pi\Delta})+\ln(\sin\frac{\pi\Delta}{4})+\frac{1}{2}\ln(\cos\frac{\pi\Delta}{4})+\frac{3}{4}\ln 2+\frac{i}{4\pi}\Big(\frac{\pi^{2}\Delta^{2}}{4}+Li_{2}(e^{i\pi\Delta})\Big)+C_{2}\,.
\ee
Here $C_{2}$ is an integration constant. 
\subsection{ Product representation}
In this section we express the free energy obtained above as an infinite product. This will be useful for comparison with the answers obtained by Green's function as well as index method. We follow the strategy that the free energy satisfies a first order differential equation, such as \eqref{eq:AdS2;Free,diff,R<1} and \eqref{eq:AdS2:L=2.2}, and identify the differential equation with a differential equation satisfies by certain infinite product. We find that these infinite products are combinations of certain double sine functions \cite{Kurokawa:2003}.
\subsection*{$L=1$}
{\bf Case : $0\leq \Delta<1$}\\
Let us consider the following function
\be
\wt S_{2}(z)=\frac{\prod_{n,p=0}^{\infty}(n+p+1-z)}{\prod_{n,p=0}^{\infty}(n+p+1+z)}\,.
\ee
We see that
\be
\frac{\wt S'_{2}(z)}{\wt S_{2}(z)}=-\lim_{s\rightarrow 1}\Big[\tilde\zeta(s,1-z)+\tilde\zeta(s,1+z)\Big]\,,
\ee
where
\be\label{eq:HurwitZ.1}
\tilde\zeta(s,a)=(1-a)\zeta(s,a)+\zeta(s-1,a)\,,
\ee
and $\zeta(s,a)$ is the Hurwitz zeta function. Taking into account the following relations
\bea\label{eq:HurwitZ.2}
&&\zeta(s,a)=\frac{1}{s-1}-\psi(a)+\mathcal O(s-1),\,\quad\zeta(0,a)=\frac{1}{2}-a \,.
\eea
Here $\psi(a)=\frac{\Gamma'(a)}{\Gamma(a)}$ which satisfies following relations
\be\label{eq:Psi.1}
\psi(x+1)=\psi(x)+\frac{1}{x},\quad \psi(1-x)-\psi(x)=\pi\,\cot\pi x\,.
\ee
We get
\be\label{eq:AdS2:InfintProd.S.1.1}
\frac{\wt S'_{2}(z)}{\wt S_{2}(z)}=\pi z\cot\pi z\,.
\ee
Thus comparing with \eqref{eq:AdS2;Free,diff,R<1}, we see that $F(\Delta)=\ln \tilde S_{2}(\frac{\Delta}{2})-\ln A'$, where $A'$ is independent of $\Delta$. Thus the partition function is
\be \label{lequal1a}
Z=e^{-F}=A'\frac{\prod_{n,p=0}^{\infty}(n+p+1+\frac{\Delta}{2})}{\prod_{n,p=0}^{\infty}(n+p+1-\frac{\Delta}{2})}=A'\frac{\prod_{r=1}^{\infty}(r+\frac{\Delta}{2})^{r}}{\prod_{r=1}^{\infty}(r-\frac{\Delta}{2})^{r}}\,.
\ee
{\bf Case : $1\leq \Delta<2$}\\
Let us consider the following function
\be
S_{2}(z)=\frac{\prod_{n,p=0}^{\infty}(n+p+z)}{\prod_{n,p=0}^{\infty}(n+p+2-z)}\,.
\ee
We see that
\be
\frac{S'_{2}(z)}{S_{2}(z)}=\lim_{s\rightarrow 1}\Big[\tilde\zeta(s,z)+\tilde\zeta(s,2-z)\Big]\,.
\ee
Again using the relations \eqref{eq:HurwitZ.1}, \eqref{eq:HurwitZ.2} and \eqref{eq:Psi.1}, we get
\bea\label{eq:AdS2:InfintProd.S.1.2}
\frac{S'_{2}(z)}{S_{2}(z)}=(1-z)[\psi(2-z)-\psi(z)]+\frac{1}{2}-z+\frac{1}{2}-(2-z)=(1-z)\pi\cot\pi z\,.
\eea
Thus comparing with \eqref{eq:AdS2;Free,diff,R<2} the natural answer for the free energy for this range of the R-charge is
\be
F(\Delta)=\ln S_{2}(\frac{\Delta}{2})-\ln \wt A\,,
\ee
and the partition function is
\be \label{lequal1b}
Z=e^{-F}=\frac{\wt A}{S_{2}(\frac{\Delta}{2})}=\wt A\,\frac{\prod_{n,p=0}^{\infty}(n+p+2-\frac{\Delta}{2})}{\prod_{n,p=0}^{\infty}(n+p+\frac{\Delta}{2})}=\wt A\,\frac{\prod_{r=1}^{\infty}(r+1-\frac{\Delta}{2})^{r}}{\prod_{r=1}^{\infty}(r-1+\frac{\Delta}{2})^{r}}\,,
\ee
where $\wt A$ is some constant. 
\subsection*{$L=2$}
Let us start with the range $1\leq\Delta<2$.\\
{\bf Case : $1\leq\Delta<2$}\\
We begin with the expression
\be
S_{2}(x,(1,2))=\frac{\Gamma(3-x,(1,2))}{\Gamma(x,(1,2))}\,,
\ee
where
\be
\Gamma(x,(w_{1},w_{2}))=\frac{1}{\prod^{\infty}_{n_{1},n_{2}=0}(n_{1}w_{1}+n_{2}w_{2}+x)}\,.
\ee
Thus in the form of infinite product we have
\be\label{defs2}
S_{2}(x,(1,2))=\frac{\prod^{\infty}_{n_{1},n_{2}=0}(2n_{1}+n_{2}+x)}{\prod^{\infty}_{n_{1},n_{2}=0}(2n_{1}+n_{2}+3-x)}\,.
\ee
We also see that the function $S_{2}(x,(1,2))$ can be written as
\be
S_{2}(x,(1,2))=S_{2}\Big(\frac{x}{2}\Big)\,S_{2}\Big(\frac{x+1}{2}\Big)\,.
\ee
Differentiating $S_{2}(x,(1,2))$ with respect to $x$, we get
\bea\label{eq:AdS2:InfintProd.S.2.1}
\frac{S'_{2}(x,(1,2))}{S_{2}(x,(1,2))}&=&\frac{1}{2}\frac{S'_{2}(\frac{x}{2})}{S_{2}(\frac{x}{2})}+\frac{1}{2}\frac{S'_{2}(\frac{x+1}{2})}{S_{2}(\frac{x+1}{2})}\,,\nn\\&=&\frac{1}{2}\Big(1-\frac{x}{2}\Big)\pi\cot\Big(\frac{\pi x}{2}\Big)+\frac{1}{2}\Big(1-\frac{x+1}{2}\Big)\pi\cot\Big(\pi\frac{ x+1}{2}\Big)\,,\nn\\&=&\frac{\pi}{4}\Big[(2-x)\cot\Big(\frac{\pi x}{2}\Big)-(1-x)\tan\Big(\frac{\pi x}{2}\Big)\Big]\nn\\&=&\frac{\pi}{4\sin\pi x}\Big[(3-2x)\cos\pi x+1\Big]\,.
\eea
Comparing with \eqref{eq:AdS2:L=2.2} we see that in this case we have
\be
\frac{dF}{dx}=\frac{S'_{2}(x,(1,2))}{S_{2}(x,(1,2))}\,,\qquad \text{for}\,\, x=\frac{\Delta}{2}\,,
\ee
which means $F=\ln S_{2}(x,(1,2))+\ln A $, therefore
\be
Z=\frac{A}{S_{2}(x,(1,2))}\Big|_{x=\frac{\Delta}{2}}\,, 
\ee
where $A$ is some constant.
Therefore using (\ref{defs2}),   upto an $\Delta$ independent constant we obtain 
\begin{equation} \label{l2a}
\ln Z =  \sum_{n_1, n_2 =0}^\infty \left( 
\ln ( 2n_1 + n_2  + 3 - \frac{\Delta}{2} )  - \ln ( 2n_1 + n_2 +  \frac{\Delta}{2}  \right)\,.
\end{equation}
{\bf Case : $0\leq\Delta<1$}\\
We notice that the differential equation \eqref{eq:AdS2:L=2.1} can be written as
\be
\frac{dF}{dx}=\frac{\pi}{4\sin\pi x}\Big[1+(3-2x)\cot\pi x\Big]+\pi(2x-1)\cot\pi x\,,
\ee
where again $x=\frac{\Delta}{2}$. We see that the first part of the RHS in the above equation is same as the RHS of \eqref{eq:AdS2:InfintProd.S.2.1}. The last term can be written as the linear combination of \eqref{eq:AdS2:InfintProd.S.1.1} and \eqref{eq:AdS2:InfintProd.S.1.2}.
\be
\frac{dF}{dx}=\frac{S'_{2}(x,(1,2))}{S_{2}(x,(1,2))}+\frac{\wt S'_{2}(x)}{\wt S_{2}(x)}-\frac{S'_{2}(x)}{S_{2}(x)}\,.
\ee
Thus the partition function is given as
\be
Z=\frac{\hat A\,S_{2}(x)}{S_{2}(x,(1,2))\wt S_{2}(x)}\Big|_{x=\frac{\Delta}{2}}\,.
\ee
Here $\hat A$ is constant.  Using the definitions of the  functions 
$S$ and $\tilde S$, the free energy up to a constant independent of $\Delta$ is given by 
\begin{eqnarray}\label{l2b}
\ln Z &=&  \sum_{n_1=1, n_2 =0}^\infty  2\ln ( n_1 + n_2 + \frac{\Delta}{2} )
+ \sum_{n_1=0}^\infty \ln ( n_1 + \frac{\Delta}{2} )  \\ \nonumber
&& -  \sum_{n_1, n_2=1}^\infty 2
\ln ( n_1 + n_2 - \frac{\Delta}{2} ) - \sum_{n_1 = 1}^\infty \ln ( n_1 - \frac{\Delta}{2} ) \\
\nonumber
& &  + \sum_{n_1, n_2 =0}^\infty \left( 
\ln ( 2n_1 + n_2  + 3 - \frac{\Delta}{2} )  - \ln ( 2n_1 + n_2 +  \frac{\Delta}{2}  \right)\,.
\end{eqnarray}

\section{Integrals involving product of hypergeometric functions} \label{appendixc}

The integrals necessary for evaluating the bulk contribution to the 
one loop determinant of the standard action  when  there exists an 
integer in $D$ involve products of hypergeometric functions. 
They are given by
\bea
&&\int^{1}_{0}dz\,\frac{ L q (-i + 2 L q \a) S_{1+}(z) S_{2-}(z)}{4c_{1+-} \sqrt{1- z}}=\frac{iLq}{2}\Big[\psi\Big(\frac{1}{2}+\frac{1}{4}x\Big)-\psi\Big(\frac{1}{4} x^{*}\Big)\Big]\,,\nn\\
&&\int^{1}_{0}dz\,\frac{ L q (-i + 2 L q \a) S_{1+}(z) S_{2+}(z)}{4c_{1++} \sqrt{1- z}}=-\frac{iLq}{2}\Big[\psi\Big(\frac{1}{2}+\frac{1}{4}y\Big)-\psi\Big(1+\frac{1}{4} y^{*}\Big)\Big]\,,\nn\\
&&\int^{1}_{0}dz\,\frac{ L q (-i + 2 L q \a) S_{1-}(z) S_{2-}(z)}{4c_{1--} \sqrt{1- z}}=\frac{iLq}{2}\Big[\psi\Big(\frac{1}{2}+\frac{1}{4}\wt x\Big)-\psi\Big(\frac{1}{4} {\wt x}^{*}\Big)\Big]\,,\nn\\
&&\int^{1}_{0}dz\,\frac{ L q (-i + 2 L q \a) S_{1-}(z) S_{2+}(z)}{4c_{1-+} \sqrt{1- z}}=-\frac{iLq}{2}\Big[\psi\Big(\frac{1}{2}+\frac{1}{4}\wt y\Big)-\psi\Big(1+\frac{1}{4} {\wt y}^{*}\Big)\Big]\,.
\eea
where $\psi(z)$ is a digamma function  defined by 
\begin{equation}
\psi(z) =  \frac{d}{dz} \ln ( \Gamma(z))\,,
\end{equation}
and $x$ and $y$ are given as
\bea
x=2p+\Delta-2Ln+2iLq\a,\quad y=2p-\Delta+2Ln-2iLq\a\,,
\eea
and $\wt x$ and $\wt y$ are obtained from $x$ and $y$ by replacing $p\rightarrow -p$, respectively. Finally, the ${}^{*}$ on $x,y,\wt x$ and $\wt y$ denotes the complex conjugation. 

The integrals necessary for evaluating the bulk contribution to the 
one loop determinant of $Q$-exact action  when  there exists an 
integer in $D$ involve products of hypergeometric functions. 
They are given by
\bea
&&\int^{1}_{0}dz\,\frac{ L^{2} q^{2}\a  \wt S_{1+}(z) \wt S_{2-}(z)}{2\wt c_{1+-} \sqrt{1- z}}=\frac{L^{2}q^{2}\a(\psi(\wt a_{2})-\psi(\wt b_{2}))}{2(\wt b_{2}-\wt a_{2})}\,,\nn\\
&&\int^{1}_{0}dz\,\frac{ L^{2} q^{2} \a\wt S_{1+}(z) \wt S_{2+}(z)}{2\wt c_{1++} \sqrt{1- z}}=\frac{L^{2}q^{2}\a(\psi(a_{2})-\psi(b_{2}))}{2(b_{2}-a_{2})}\,\nn\\
&&\int^{1}_{0}dz\,\frac{ L^{2} q^{2}\a  \wt S_{1-}(z) \wt S_{2-}(z)}{2\wt c_{1--} \sqrt{1- z}}=\frac{L^{2}q^{2}\a(\psi(\wt a_{2}-p)-\psi(\wt b_{2}-p))}{2(\wt b_{2}-\wt a_{2})}\,,\nn\\
&&\int^{1}_{0}dz\,\frac{ L^{2} q^{2} \a\wt S_{1-}(z) \wt S_{2+}(z)}{2\wt c_{1-+} \sqrt{1- z}}=\frac{L^{2}q^{2}\a(\psi(a_{2}-p)-\psi(b_{2}-p))}{2(b_{2}-a_{2})}\,,
\eea
where the arguments of the digamma function are given in \eqref{eq:AdS2.HyperCoeff}.

\providecommand{\href}[2]{#2}\begingroup\raggedright\endgroup


\end{document}